\documentclass[twocolumn,times]{aastex631}

\newcommand*{\msun}{\ensuremath{\mathrm{M}_{\odot}}}
\newcommand*    \kms{{\,\rm km\,s^{-1}}}
\newcommand*    \pc{{\,\mathrm{pc}}}
\newcommand*    \kpc{{\,\mathrm{kpc}}}

\newcommand*    \mbh{M_\bullet}
\newcommand*    \af{a_f}
\newcommand*    \ah{a_h}
\newcommand*    \tk{t_k}
\newcommand*    \vk{v_k}
\newcommand*    \ve{v_{\mathrm{esc}}}

\newcommand*    \re{r_e}
\newcommand*    \fN{f_N}
\newcommand*    \rb{r_b}
\newcommand*    \rk{r_k}
\newcommand*    \ri{r_i}
\newcommand*    \rn{r_N}
\newcommand*    \en{\epsilon_N}
\newcommand*    \tbr{t_\mathrm{Br}}
\newcommand*    \mn{M_N}
\newcommand*    \sn{\sigma_N}
\newcommand*    \fn{f_N}
\newcommand*    \mb{M_b}
\newcommand*    \fhr{f_{\mathrm{hr}}}
\newcommand*    \sersic{S\'{e}rsic }
\newcommand*    \cs{core-S\'{e}rsic }

\defcitealias{khonji2024core}{Paper~1}

\begin{document}

\title{Black hole dragging: a new mechanism for forming nuclear star clusters in giant elliptical galaxies}

\shorttitle{NSCs from Black Hole Dragging}
\shortauthors{Khonji et al.}

\correspondingauthor{Nader Khonji}
\email{n.khonji@surrey.ac.uk}

\author[0000-0002-6633-2185]{Nader Khonji}
\affiliation{School of Mathematics and Physics, University of Surrey, Guildford GU2 7XH, UK}

\author[0000-0002-9420-2679]{Alessia Gualandris}
\affiliation{School of Mathematics and Physics, University of Surrey, Guildford GU2 7XH, UK}

\author[0000-0002-1164-9302]{Justin I. Read}
\affiliation{School of Mathematics and Physics, University of Surrey, Guildford GU2 7XH, UK}

\author[0000-0001-8669-2316]{Walter Dehnen}
\affiliation{Astronomisches Rechen-Institut, Zentrum f{\"u}r Astronomie der Universit{\"a}t Heidelberg, M{\"o}nchhofstra\ss{}e 12-14, Heidelberg 69120, Germany}

\begin{abstract}
It has long been thought that nuclear star clusters (NSCs) cannot co-exist with the most massive supermassive black holes (SMBHs), since SMBH mergers – unavoidable for the most massive systems – would scatter away NSC stars. However, central concentrations of light have now been reported in up to a third of all massive ellipticals. We present a new mechanism for forming NSCs in giant elliptical galaxies, arising naturally from SMBH mergers, which could explain these observations. We call this “black hole dragging”. After a major merger of two galaxies and their SMBHs, the newly-merged SMBH can receive a gravitational wave recoil kick. We show that recoiling SMBHs induce two competing effects on the galaxy’s background stars. Firstly, some stars become bound to the SMBH and co-move with it, an effect strongest at low recoil velocities. Secondly, background stars are ejected as the recoiling SMBH falls back due to dynamical friction, an effect strongest at high recoil velocities. At intermediate recoil velocities (500\mbox{--}1000 km/s), both effects become important, and the density of bound stars can exceed that of the background stellar core. This yields a central dense NSC that is clearly visible in the galaxy’s surface brightness profile. We show that NSCs formed in this way have realistic sizes, masses and velocity dispersions when measured similarly to observed systems. This provides a route for even giant ellipticals containing SMBHs to host an NSC. We predict such NSCs should have indistinguishable colors, ages and chemistry from non-NSC central stars, combined with low ellipticities.
\end{abstract}

\section{Introduction} \label{sec:intro}
Nuclear star clusters (NSCs) are dense, bright stellar clusters, located in the central regions of galaxies. They have been defined as ``stellar light above the inward extrapolation of the host galaxy's surface brightness profile on scales of $\lesssim 50 \pc$" \citep{neumayer2020nuclear}, and usually coincide with the photometric \citep{boker2002hubble} and dynamical \citep{neumayer2011two} center of their host galaxy. However, there is a lack of clarity in the terminology, and the same objects can often be referred to as ``nuclei". \citep [e.g.][]{lauer2005centers}.

Most NSCs are small, with a median effective radius ($\re$) of $3 \pc$ \citep{neumayer2020nuclear}, similar to globular clusters \citep[GCs,][]{harris2010new}, but the distribution has a significant tail towards large $\re$
, with NSCs in some spiral and elliptical galaxies reaching $\sim 50\pc$ \citep{georgiev2016masses, cote2006acs}. NSC stellar masses ($\mn$) are much larger than those of GCs and scale with galaxy stellar mass  ($M_*$), approximately as $\mn \propto M_*^{1/2}$ \citep{balcells2002galactic, scott2013updated}, though the relation appears almost linear for high mass NSCs in spiral galaxies \citep{neumayer2020nuclear}. Many NSCs are non-spherical, with ellipticities of up to 0.6 \citep{neumayer2020nuclear}. In elliptical galaxies there is a correlation between the ellipticity and mass of their NSCs \citep{spengler2017virgo}.

The determination of stellar ages and metallicities is extremely difficult for all but the nearest galaxies, so population fitting of integrated light must be used \citep{tinsley1978evolutionary} to determine the most likely age-metallicity combination. For elliptical galaxies, this indicates that metallicities are generally lower in low-mass galaxies than in those of higher mass, both in absolute terms and relative to that of their host galaxies \citep{neumayer2020nuclear}. However, there is significant stochasticity for individual galaxies. In spiral galaxies, spectroscopic studies indicate a mixture of stellar populations. Although old stars dominate in terms of mass, there is a significant young population. 

There are two main theories for the formation of NSCs: (i) dynamical friction followed by merging of GCs \citep{tremaine1975formation, lotz2001dynamical}; and (ii) in-situ star formation after infall of gas \citep{loose1982bursts, mihos1996gasdynamics}.  A combination of these two is also possible and can occur naturally in galaxy mergers, especially if star formation prior to the merger is suppressed \citep [e.g.][]{gray2024edge}.

Dynamical friction \citep{chandrasekhar1943dynamical} of GCs was first proposed by \cite{tremaine1975formation}, shortly after the detection of a NSC in M31 \citep{light1974nucleus}. Their calculations show that a NSC of $\mn\sim10^7\,\msun$ could form in $\sim10^{10}$ years. Observational evidence for this mechanism includes a deficit of GCs in the central regions \citep{lotz2001dynamical, capuzzo2009globular}, and a correlation between the fraction of galaxies with NSCs and those with GCs in low-mass cluster ellipticals observed in the Virgo cluster \citep{sanchez2019next}. Furthermore, semi-analytic simulations have formed NSCs with density profiles \citep{antonini2013origin}, and mass and radius \citep{gnedin2014co}, that are largely consistent with observations. In addition, $N$-body simulations have matched their shape \citep{capuzzo2008self,hartmann2011constraining, antonini2012dissipationless}. They also show that, if sufficient GCs are formed early enough, NSCs could be produced from dynamical friction of GC systems within a Hubble time. However, this mechanism does not explain the significant proportion of young stars seen in NSCs of spiral galaxies.

Infall of gas to the nucleus generally leads to star formation, a process that may occur more than once if interrupted by gas dispersion through supernova feedback and stellar winds followed by later gas infall \citep{loose1982bursts}. Wet mergers are likely to lead to gas infall and starbursts \citep{mihos1996gasdynamics, gray2024edge}. However, since spiral galaxies which show no signs of previous mergers also have NSCs, other mechanisms have been proposed. These include bar-driven inspiral \citep{shlosman1990fuelling,schinnerer2006molecular} and mergers of gaseous spiral arms \citep{bekki2007formation}.

\cite{neumayer2020nuclear} argue that GC dynamical friction and gas infall are more important in lower and higher mass galaxies respectively, with the transition occurring at $M_*\sim 10^9\;\msun$, but with significant scatter. Their evidence for this includes the metallicity differences described above and the finding that the fraction $\fn$ of galaxies with NSCs tracks the fraction of galaxies with GCs \citep{sanchez2019next}.

Imaging via the Hubble Space Telescope finds $\fN$ to be $\sim50\%$ for spiral galaxies \citep{carollo1997spiral, carollo2002spiral}, increasing to $\sim80\%$ for late-type spirals \citep{boker2002hubble}. Surveys of galaxy clusters show that most elliptical galaxies also have detectable NSCs: $\fN=66$\mbox{--}82\% for elliptical galaxies in the Virgo cluster \citep{cote2006acs} and $\fN\sim80\%$ for dwarf elliptical galaxies in the Coma cluster \citep{den2014hst}. 

In general, $\fn$ varies with $M_*$ (or galaxy luminosity). Overall, there are nuclei in the vast majority of galaxies of intermediate stellar mass ($M_*=10^{8\text{-}10}\,\msun$). There is relatively little data available for spiral galaxies, but what there is suggests that $\fn$ drops at low brightness, to around $0.1$ at $M_V > -17$ \citep{georgiev2009globular}. 

For elliptical galaxies, $\fn$ peaks at $> 0.9$ for galaxies with $M_* \sim 10^9 \, \msun$ \citep{sanchez2019next} or $M_I \sim -18$ \citep{den2014hst}. At the low mass / brightness end it falls to around zero for $M_* \sim 10^5 \, \msun$ \citep{sanchez2019next} or $M_I \sim -10$ \citep{munoz2015unveiling}, which could be partly owed to the difficulty of detecting faint nuclei. There is also a drop in $\fn$ above $10^9 \msun$. \cite{cote2006acs} found $\fn\sim 0$ for $M_B \lesssim -20.5$ and related it to the transition from “cusp" to “core" galaxies; the surface brightness profiles of these brighter elliptical galaxies tends to be flattened, forming a central core, whereas those of dimmer galaxies tend to continue to rise steeply as a cusp. 

Central constant luminosity cores at the centers of some giant elliptical galaxies are thought to form as the result of dry mergers between galaxies containing supermassive black holes (SMBHs). Since NSCs were found to have similar relations with their host galaxies to SMBHs \citep{rossa2006hubble}, it was initially thought that NSCs were the low-mass counterparts of SMBHs. It also seemed likely that if a SMBH binary formed after a merger, the transfer of energy and angular momentum from the binary would destroy the NSC \citep{quinlan1997dynamical, milosavljevic2001formation}. It is now clear that SMBHs and NSCs can coexist, not least in the case of the Milky Way \citep{schodel2009nuclear}, and there is evidence that NSCs do exist in giant elliptical galaxies. \cite{lauer2005centers} found nuclei in more than a third in their survey of 77 massive ellipticals, and $20\%$ of the 40 galaxies with available colors have nuclei with similar absorption line spectra to the main galaxy \citep{neumayer2020nuclear}.

\begin{figure}
    \centering
    \includegraphics[width = \columnwidth]{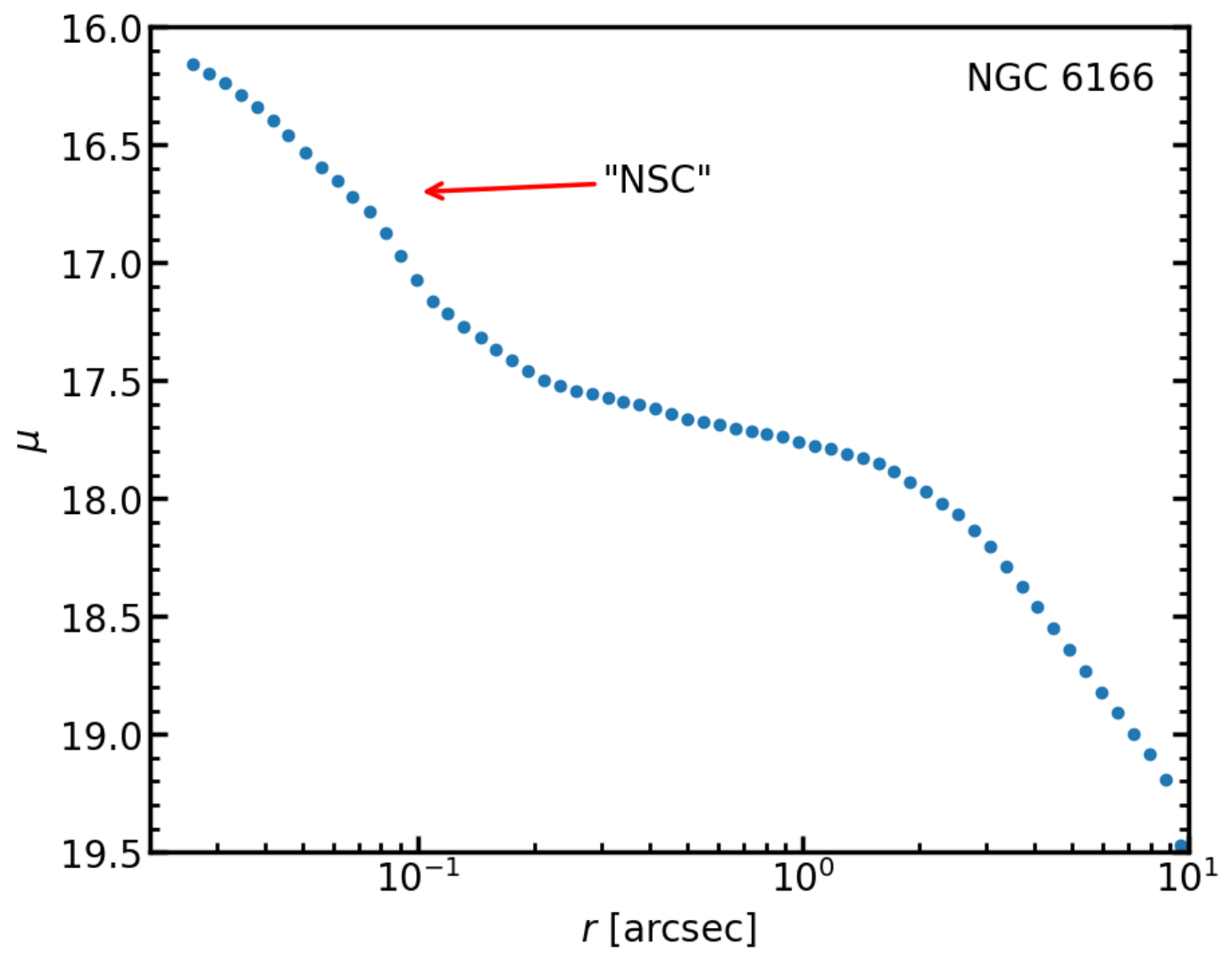}
    \caption{Surface brightness profile of NGC 6166. Data obtained from \cite{dullo2019most}.}
    \label{fig:ngc6166_sb}
\end{figure}

Similarly, \cite{dullo2019most} found ``additional nuclear light components" in $3/12$ galaxies with some of the largest known cores, and stated that they could be ``nuclear star clusters". The surface brightness profile of one of these galaxies (NGC 6166) is shown in Figure \ref{fig:ngc6166_sb}. The typical central velocity dispersion $\sigma$ of lower-mass NSCs in galaxies such as the Milky Way is $\sim60 \kms$. However, in NGC 6166 it reaches some $300 \kms$ \citep{bender2015structure}, challenging rule-of-thumb assumptions about NSC properties. It appears that central light excesses in giant cored elliptical galaxies exist. These look like NSCs in their surface brightness profiles (Figure \ref{fig:ngc6166_sb}), but with atypically large size and velocity dispersion. In this paper, we show how NSCs like this can form naturally in galaxy mergers as a result of a new mechanism, which we call ``black hole (BH) dragging."

Mergers leading to core formation can be divided into three stages \citep{begelman1980massive}. First, dynamical friction of the SMBHs against stars and dark matter (DM) slows the SMBHs until they form a gravitationally bound binary \citep{antonini2011dynamical,valtaoja1989binary}. Second, three-body interactions between the binary and stars (or dark matter) leads to stellar ejections and hardening of the binary \citep{quinlan1996dynamical, sesana2006interaction}. This second phase is known as binary scouring, and its dynamical heating is both the most accepted mechanism for core formation \citep{quinlan1996dynamical} and the proposed cause of erasure of NSCs from high mass ellipticals \citep{bekki2010transition, antonini2015coevolution}. In the final merger phase, gravitational wave (GW) emission leads to inspiral and coalescence of the SMBH binary \citep{enoki2004gravitational, sesana2013gravitational}. 

At the time of SMBH coalescence, depending on their mass ratio and spins (both magnitude and orientation), asymmetric emission of GWs may occur. Since GWs carry linear momentum, the SMBH remnant receives a ``recoil kick".  Most kicks are small: for non-spinning SMBHs, \cite{gonzalez2007maximum} find a maximum kick speed of $\vk=175\,\kms$, while for spinning SMBHs $\vk \lesssim 600 \kms$ \citep{schnittman2007distribution}. However, certain configurations of the spins can lead to very high recoil speeds. Using numerical relativity, \cite{campanelli2007maximum} show that for spins aligned and anti-aligned with the orbital angular momentum, recoil speeds of $\sim 4000 \kms$ can be achieved. Kicks of this size would eject the SMBH remnant from the galaxy but, after more modest kicks, the remnant and core are expected to oscillate about their common center of mass (COM) until they reach thermal equilibrium \citep{gualandris2008ejection}.

Although most cores are relatively small at tens to a few hundreds of $\pc$ in size, a few are $> 0.5 \kpc$ \citep{dullo2019most}. \cite{nasim2021formation} performed equal mass merger simulations based on the observed parameters of A2261-BCG, followed by GW recoil kicks, and found that scouring alone produced a core size of $\sim$ 1 kpc, but that a  GW kick of $\sim0.8 \, \ve$ could reach the observed size. In addition, they found that a mild cusp was retained after scouring, but GW recoil resulted in a flat profile. \citealt{khonji2024core} (hereafter \citetalias{khonji2024core}) performed a similar study with a range of initial conditions, based on four galaxies from \cite{dullo2019most} with observed core sizes ranging from $0.65$\mbox{--}$2.71 \kpc$, including A2261-BCG. In these simulations, after hardening of the SMBH binary, we merged the SMBHs and gave the remnant recoil kicks of between $0.5\mbox{--}0.9$ of the escape speed ($\ve$) of the galaxy remnant. We showed that binary scouring alone could form cores $< 1.3 \kpc$ in size, but that GW recoil speeds of $\lesssim 0.5 \,\ve$ are required to form cores $> 2 \kpc$. We also found that a unique signature of GW recoil heating is a truly flat core in the spatial density, which appears as a core in the surface brightness. NGC 1600 was one of the galaxies which did not require recoil, consistent with previous modeling by \cite{rantala2018formation}. Recently, \cite{rawlings2025identifying} studied GW recoil in a version of this model, scaled-down by a factor of $\sim3$ in mass, and found the core after scouring could be enlarged by a factor of $2$\mbox{--}$3$. 

In this paper, we study the effect of the likely more common smaller GW recoil kicks on the nuclei in six galaxies, four of which were those in \citetalias{khonji2024core}. We examine whether the effect of recoil heating is the same, resulting in cores that are larger and flatter than after scouring alone. Somewhat unexpectedly, we find that smaller kicks can result in central cusps in surface brightness, steeper than the profiles after binary scouring. These cusps sometimes appear within a larger flattened core, and, as such, would be classified observationally as NSCs. For this reason, we refer to these central cusps throughout this paper as “NSCs”. We show that NSC formation in our simulations owes to two competing effects. Firstly, some stars become bound to the SMBH and co-move with it, an effect that is maximized at low recoil velocity. Secondly, background stars are ejected from the galaxy center as the recoiling SMBH falls back due to dynamical friction. This effect is maximized at high recoil velocity. At intermediate recoil velocities, both effects become important and the density of bound stars can exceed that of the background stellar core. This yields a central dense NSC that forms via a new mechanism: black hole dragging.

This paper is organized as follows. The merger and GW recoil simulation methodology are described in Section~\ref{sec:methods}. The evolution of the density profiles, core fitting, formation mechanism and properties of the NSCs are in \mbox{Section \ref{sec:results}}. Finally, discussion and comparison with observed NSCs are presented in Section \ref{sec:discussion}. 

\begin{deluxetable*}{lcccccccc}
\tablecaption{Parameters of the selected galaxies.}
\label{tab:observed}      
\tablehead{
\colhead{Galaxy} & \colhead{Type} & \colhead{$\rb$} & \colhead{$\gamma$} & \colhead{$\alpha$} & \colhead{$n$}  & \colhead{$\re$} & \colhead{$M_\bullet$}\hspace{-0.2cm}\tablenotemark{\scriptsize{a}} & \colhead{$M_*$}\hspace{-0.2cm}\tablenotemark{\scriptsize{b}}\\ 
&           & \colhead{(kpc)}    &            &           &       & \colhead{(kpc)} & \colhead{($10^{10} M_\odot$)}   & \colhead{($10^{12} M_\odot$)}  }
\startdata
NGC 1600    & Isolated  & 0.65      & 0.04      & 2         & 6.3   & 22.8  & 1.70  &   1.51               \\
A2147-BCG   & BCG       & 1.28      & 0.14      & 2         & 6.4   & 31.8  & 2.63  &   1.34               \\
NGC 6166    & BCG       & 2.11      & 0.14      & 2         & 9.0   & 83.1  & 4.79  &   3.39               \\
4C +74.13   & BCG       & 2.24      & 0.28      & 2         & 3.7   & 20.9  & 5.13  &   2.39               \\      
A2261-BCG   & BCG       & 2.71      & 0.00      & 5         & 2.1   & 17.6  & 6.45  &   4.07               \\
IC 1101     & BCG       & 4.2       & 0.08      & 2         & 5.6   & 11.6  & 11.0  &   1.10                \\
\enddata
\tablerefs{\cite{dullo2017remarkably, dullo2019most}}
\tablecomments{The first five parameters are from core-S\'ersic fits (equation~\ref{eq:core-Sersic}).}\vspace{-0.2cm}
\tablenotetext{a}{Black hole masses from the $M_\bullet$-$\rb$ relation by \cite{dullo2019most}, except NGC\,1600 (directly measured)}\vspace{-0.2cm}
\tablenotetext{b}{Bulge mass from the $M_*$-$L_*$ relation by \cite{worthey1994comprehensive}.}
\end{deluxetable*}
\begin{deluxetable*}{lcccccccccc}
\tablecaption{Parameters of the $N$-body models for the precursor galaxies.}             
\label{tab:precursors}              
\tablehead{
\colhead{Galaxy} & \colhead{$M_\bullet$} & \colhead{$M_*$} & \colhead{$M_{\mathrm{dm}}$} & \colhead{Scale} & \colhead{Cutoff} & \colhead{$n$} & \colhead{$\re$} & \colhead{$N_*$} & \colhead{$N_{\mathrm{dm}}$} & \colhead{Minimum}\\[-2mm]
                &                       &                   &                   & \colhead{Radius} & \colhead{Radius} &               &                  &                &                             & \colhead{Particle Mass}          \\
                        & \colhead{($10^{10} M_\odot$)} & \colhead{($10^{12} M_\odot$)} &  \colhead{($10^{14} M_\odot$)} & \colhead{(kpc)} & \colhead{(Mpc)} & & \colhead{(kpc)} & \colhead{($10^4$)} & \colhead{($10^5$)} & \colhead{($10^6 M_\odot$)} }
\startdata
NGC 1600    & 0.85       & 0.76    &  0.79    & 165    & 1.65      & 6.3   & 11.4               & \phantom{5}9.49  &   9.60    & 0.84\\
A2147-BCG   & 1.32       & 0.67    &  1.00    & 183    & 1.83      & 6.4   & 15.9               & \phantom{5}6.81  &   9.86    & 1.04\\
NGC 6166    & 2.40       & 1.69    &  1.00    & 183    & 1.83      & 9.0   & 41.7               & 15.34            &   9.02    & 1.17\\
4C +74.13   & 2.60       & 1.20    &  1.00    & 183    & 1.83      & 3.7   & 10.5               & 11.31            &   9.41    & 1.12\\
A2261-BCG   & 3.23       & 2.04    &  1.00    & 183    & 1.83      & 2.1   & \phantom{5}8.8     & 17.61            &   8.31    & 1.20\\
IC 1101     & 5.50       & 0.55    &  1.00    & 183    & 1.83      & 5.6   & \phantom{5}5.8     & \phantom{5}5.48  &   9.98    & 1.06\\
\enddata
\end{deluxetable*}

\section{Methods}
\label{sec:methods}
To study the effects of binary scouring and GW recoil on the nuclei of giant elliptical galaxies, we performed $N$-body simulations of equal-mass mergers, modeled on six such galaxies selected for their large core sizes of $>0.5\kpc$. Five of the chosen galaxies were taken from \cite{dullo2019most}, four of which were also used in \citetalias{khonji2024core} studying the effects of GW recoil kicks with $\vk\geq 0.5 \,\ve$. The remaining galaxy (IC 1101) was taken from \cite{dullo2017remarkably}, and is notable for its extremely large core size of $4.2\kpc$. NGC 1600 is the only galaxy in the sample with a dynamical measurement of the SMBH mass $\mbh$ \citep{thomas201617}; therefore, for the other galaxies, we used the relation provided by \cite{dullo2019most} between the SMBH mass and the core radius $\rb$, obtained by fitting a \cs profile. This combines an outer S\'ersic profile \citep{sersic1963influence} with an inner logarithmic slope $\gamma$, with intensity $I$ given by:
\begin{equation}
\label{eq:core-Sersic}
I = I' \Bigg[1+\bigg(\frac{r_\mathrm{b}}{r}\bigg)^\alpha\Bigg]^{\frac{\gamma}{\alpha}} \mathrm{exp} \,\Bigg(-b_n\bigg[\frac{r^\alpha + r_\mathrm{b}^\alpha}{r_\mathrm{e}^\alpha}\bigg]^{1/\alpha n}\Bigg) \ ,
\end{equation}
where $r_b$ is the core radius, $\alpha$ is the sharpness of transition between the inner and outer parts of the profile, $n$ is the S\'ersic index, and $b_n \approx 2n - 1/3$ ensures that $\re$ is the effective radius of the galaxy \citep{sersic1963influence}. Finally, $I'$ is defined as:
\begin{equation}
\label{eq:idash}
I'=I_\mathrm{b} \, 2^{-\gamma/\alpha}\, \mathrm{exp}\,\Bigg[b_n\bigg(\frac{2^{1/\alpha}r_\mathrm{b}}{r_\mathrm{e}}\bigg)^{1 / n}\Bigg] \ .
\end{equation} 

The bulge masses were calculated by \cite{dullo2017remarkably} and \cite{dullo2019most}, assuming an age of 12 Gyr, from their observed colors and metallicities, using the $M_*-$Luminosity relation from \cite{worthey1994comprehensive}. The relevant parameters for all six galaxies are given in Table~\ref{tab:observed}. The setup of the galaxy models and the properties of the $N$-body simulations are described in detail in \citetalias{khonji2024core}, but the key elements are summarized here. 

We simulated each galaxy as the remnant of an equal-mass merger of two identical precursor galaxies. The model parameters for the precursors are summarized in Table~\ref{tab:precursors} and were derived from the corresponding observed values in Table~\ref{tab:observed}. We assumed that the total stellar and SMBH masses were conserved during the respective dry galaxy and SMBH mergers, such that the precursor masses were half those of the observed galaxies. We further assumed that the precursors have \sersic profiles with index $n$ identical to that observed for the remnants, but with half the effective radius $\re$, as predicted by the virial theorem for dissipationless equal-mass merger \citep[e.g.][]{binney2011galactic, barnes1992transformations, naab2009minor}.

Five out of six galaxies are brightest cluster galaxies (BCGs), such that their DM halos are contiguous with those of neighboring galaxies. Hence, their enclosed mass $M_{200}$ at mean density 200 times the critical density was estimated to be $\sim10^{14} M_\odot$, using its relation with $M_*$ given by \cite{correa2020dependence}. For NGC 1600, which is unusually isolated, we used half of the estimated mass of $\sim15.8 \times 10^{14} M_\odot$ \citep{goulding2016massive} for the precursors.

We used the AGAMA library \citep{vasiliev2019agama} to create equilibrium multicomponent models (SMBH, stellar bulge and DM halo). \sersic \citep{sersic1963influence} and Navarro-Frenk-White \citep{navarro1997universal} profiles were used for the bulge and halo components, respectively. For the former, AGAMA uses a deprojected profile in which the bulge mass can be used as a proxy for luminosity. Potential models of all three components and density models for bulge and halo were used to create spherical ergodic distribution functions. These were then sampled to provide the final galaxy $N$-body models.

The total particle number $N$ was $\sim10^6$, balancing resolution and resource usage. The number of bulge ($N_*$) and halo ($N_{\mathrm{dm}}$) particles were chosen so that the halo:bulge particle mass ratio (PMR) was $\sim 10$. However, central resolution was enhanced by the use of a mass refinement scheme \citep{attard2024multiresolution}, which over-samples the central particles by a factor of 10, as follows. Bulge and halo particles were each divided into four concentric zones by radius, respectively containing  $1\%$, $11.5\%$, $38.5\%$ and $49\%$ of particles. The innermost zone was left with the full increased resolution. From the second zone outwards, particle numbers are reduced, and particle masses correspondingly increased, by factors of $\sim 2.53$, $10$ and $40$, respectively, such that $N$ is restored to the original value and the PMR is maintained in each zone.

The merger simulations were performed using the \textsf{griffin} $N$-body code that employs as gravity solver an adaptive fast multipole method with adjustable error \citep{dehnen2014fast}, which we set to an average relative force error of 0.03\%. Plummer-type softening was used, with a parameter of $3\pc$ for interactions involving the SMBHs and $30\pc$ for all other interactions. The mass shells were conservatively chosen to avoid mixing, given the fixed softening used.

The precursor galaxies were placed in a highly eccentric orbit ($e=0.95$), mimicking those in cosmological simulations \citep{khochfar2006orbital, fastidio2024}, and resulting in faster orbital evolution \citep{gualandris2022eccentricity}. The initial separation was $80\kpc$, providing adequate separation of the bulges. 

The SMBHs were followed through the dynamical friction phase, which ends with the formation of a black hole binary (BHB) at a separation $\af$, defined as the separation where the mass enclosed, from the BHB COM, is equal to the mass of the BHB: 
\begin{equation}
    \label{eq:af}
    M_* \,(r < \af) = 2\,M_\bullet \,.
\end{equation}
The BHB then undergoes rapid hardening by three-body encounters with stars. Due to the triaxiality of the merger remnant \citep{gualandris2017collisionless, bortolas2018triax}, this continues efficiently until the BHB reaches the hard binary separation:
\begin{equation}
    \label{eq:ah}
    \ah = \frac{GM_\bullet}{4\sigma^2} \ ,
    \end{equation}
where $\sigma$ is the stellar velocity dispersion. This corresponds to the distance at which the specific binding energy exceeds the specific kinetic energy \citep{milosavljevic2001formation}. After BHB hardening, the $N$-body simulations are halted due to the high computational costs (owing to the very short time steps required to resolve the dynamics close to the BHB), and a phase of GW emission and subsequent coalescence of the BHB is assumed. To this end, the BHB in each model was replaced with a single SMBH of the same total mass and COM. The merged SMBH was then given a ``GW recoil kick", arbitrarily in the $x$-direction\footnote{We note that this choice is not exactly random, as the 
orbital plane of the precursor orbit introduces a preferential direction. As a result, the $x$-direction deviates very slightly from the remnants' long axes.} with magnitude $\vk$ equal to 0.1, 0.3 or 0.5 of the escape speed $\ve$ of the remnant, calculated from the combined bulge and halo potential. The $N$-body simulations were resumed with the single SMBH.

The SMBH motion following the kick can be characterized as a damped oscillator \citep{gualandris2008ejection} with passages in and out of the galaxy center until its velocity reaches the ``Brownian velocity" $v_\mathrm{B}$, given by equipartition of kinetic energy with the stars \citep{smoluchowski1906essai, merritt2001brownian, bortolas2016brownian}: 
\begin{equation}
\label{eq:brownian}
v_{\mathrm{B}}^2 = \frac{m}{M_\bullet} \sigma^2 \ ,
\end{equation} 
where $m$ is the mass of a stellar particle. We continued the simulations until the SMBH settles to Brownian motion. This time is clear from its maximum velocity relative to the COM of the stellar component, which shows a rapid drop to below the calculated Brownian velocity.

\newpage
\section{Results}
\label{sec:results}

\subsection{Formation of the NSCs}

\begin{figure*}
    \centering
    \includegraphics[width = \textwidth]{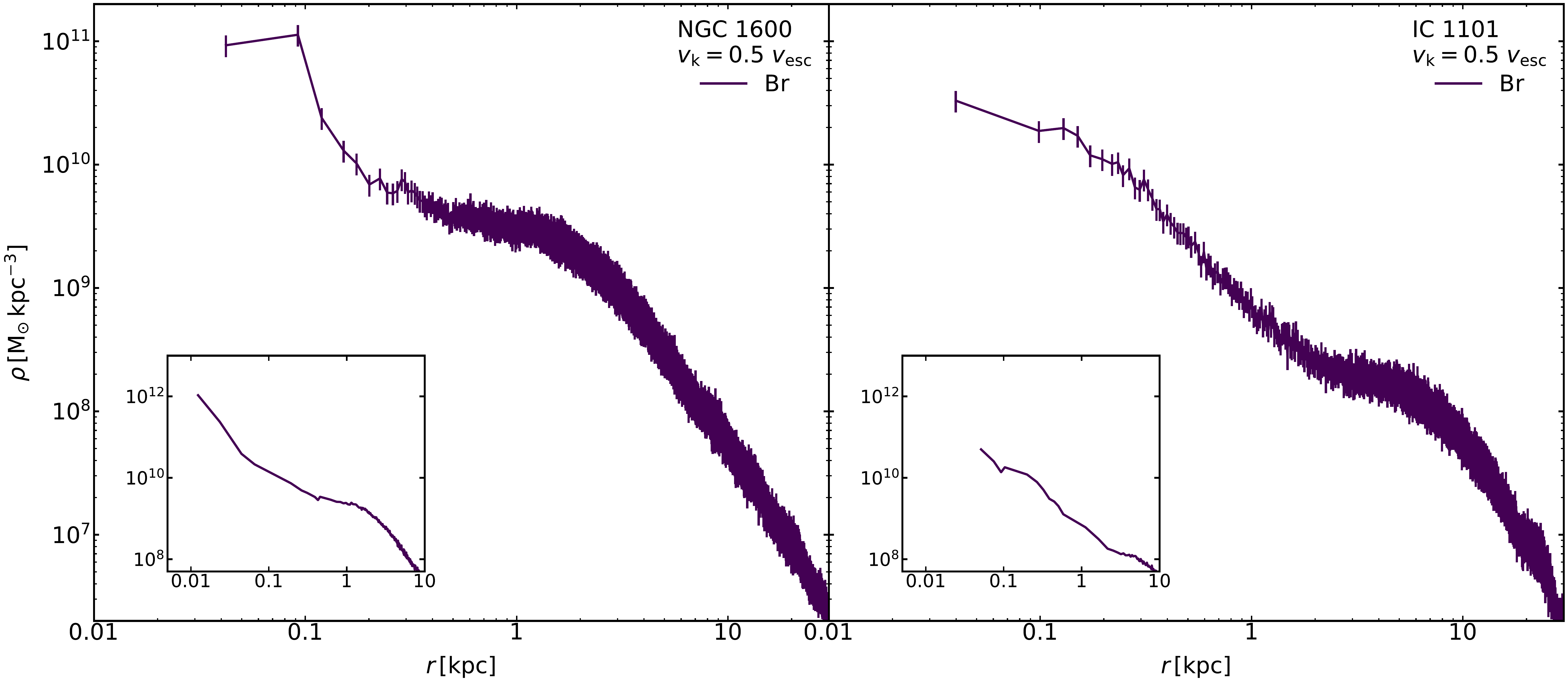}
    \caption{Volume density profiles, centered on the stellar COM, at the end of the recoil kick. This is at ``Br", which indicates the time the SMBH has settled into Brownian motion. The galaxy models shown are NGC 1600 (left) and IC 1101 (right), for GW recoil kicks of $\vk/\ve=0.5$. We use binned SPH profiles for the main profiles, but the inset plots use spherical shells centered on the SMBH. Increases in central density, which we interpret as NSCs, are clearly seen within a flattened core.}
    \label{fig:density_end}
\end{figure*}

\begin{figure*}
    \centering
    \includegraphics[width = \textwidth]{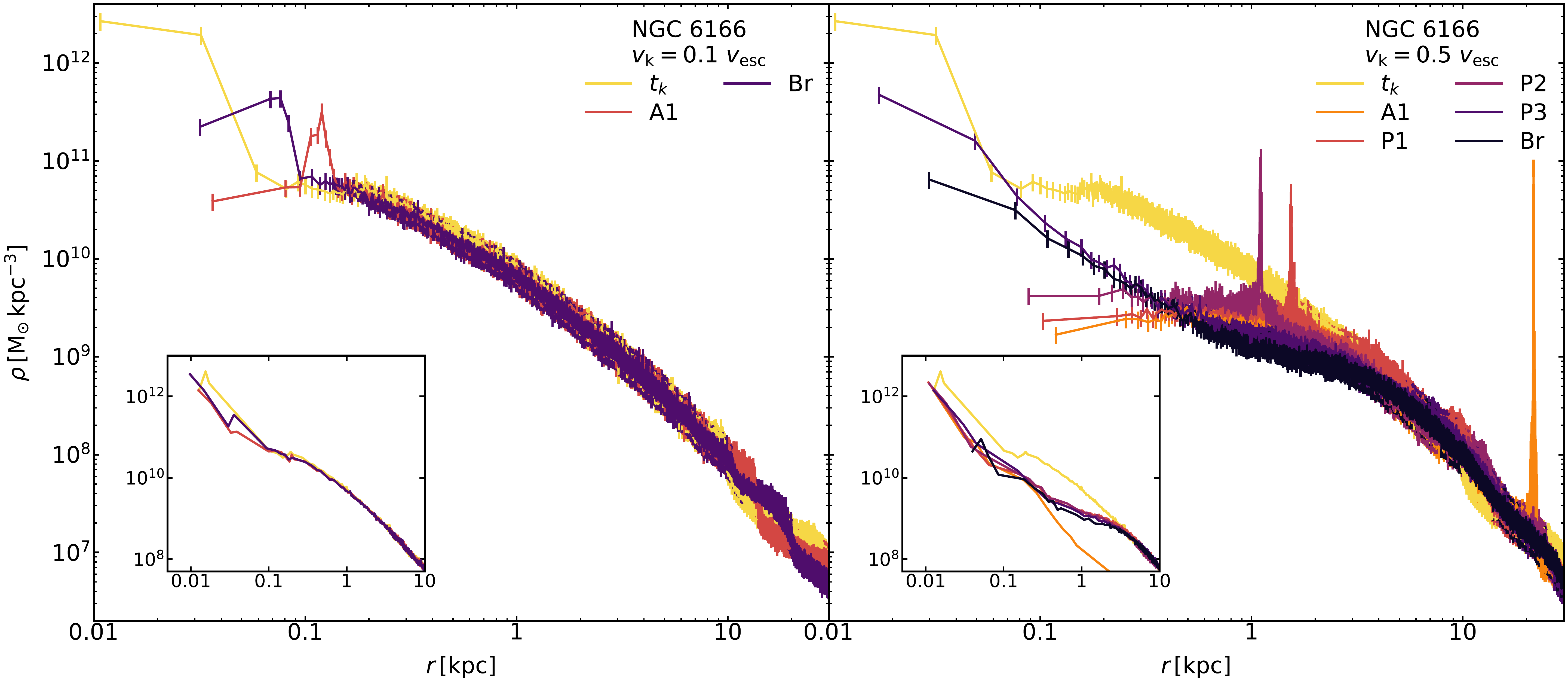}
    \caption{Evolution of the volume density profiles, centered on the stellar COM, for the $N$-body merger remnant model of NGC 6166 after GW recoil kicks of $\vk/\ve=0.1$ (left) and $0.5$ (right), using binned SPH profiles. $\tk$ is the time when the $N$-body simulation is paused for the SMBH merger and the GW recoil kick is given; ``A$n$" and ``P$n$" indicate $n$th apocenter and pericenter passages of the kicked SMBH; ``Br" indicates that the SMBH has settled into Brownian motion. The inset plots use spherical shells centered on the SMBH. For the the $\vk/\ve=0.5$ kick, the density changes as the SMBH remnant moves away from and back to the center, and the central density becomes more prominent within a flattened core.}
    \label{fig:density_evolution}
\end{figure*}

Stellar density profiles (Figures \ref{fig:density_end} and \ref{fig:density_evolution}, and Appendix \ref{sec:density}) were constructed in radial bins from individual densities estimated using an SPH-style kernel estimator \citep[using Pynbody,][]{pynbody}. This was preferred to traditional spherical shells, to better illustrate the profiles of the remnant, due to its advantages in non-spherical cases. We find that the lowest velocity kicks often have a negligible effect on the profiles (e.g. the $\vk/\ve = 0.1$ kick for NGC 6166, Figure \ref{fig:density_evolution}). In \citetalias{khonji2024core}, we showed that the highest velocity kicks result in large, flat cores. However, in intermediate cases, a central density remains within a flattened profile (Figure \ref{fig:density_end}). This is most clearly seen in A2147-BCG after a $\vk/\ve = 0.3$ kick, NGC 6166 and IC 1101 after a $\vk/\ve = 0.5$ kick, and NGC 1600 after kicks with $\vk/\ve$ of both $0.3$ and $0.5$. We interpret these increased central densities as NSCs and will focus on these five runs for the remainder of the paper. The evolution of these changes in density for NGC 6166 after $\vk/\ve = 0.5$ kick are shown in Figure \ref{fig:density_evolution}. The density profiles become flattened by the time the kicked SMBH remnant reaches its first apocenter. When the SMBH returns to the nucleus at its first pericenter, the density profiles become steeper again. After multiple passages the outer part of the core remains flattened, but the density remains increased around the SMBH in the most central region. 

To investigate the possible formation mechanism of the NSCs, we examined the trajectories of the SMBH remnants after the GW kicks (Appendix \ref{sec:traj}). For the $\vk/\ve=0.1$ kicks, the remnant travels a relatively modest distance of $\sim 1\kpc$ and is quickly damped to Brownian motion. After the $\vk/\ve=0.3$ kicks, the SMBH travels further ($\sim 3\mbox{--}7 \kpc$) but, in general, only reaches one apocenter before being damped at the first pericenter. The exception is IC 1101, which makes two passages. For the $\vk/\ve=0.5$ kicks, two or three passages are the norm, although A2147-BCG makes five, before Brownian motion is reached. 

\begin{figure*}
    \centering
    \includegraphics[width = \textwidth]{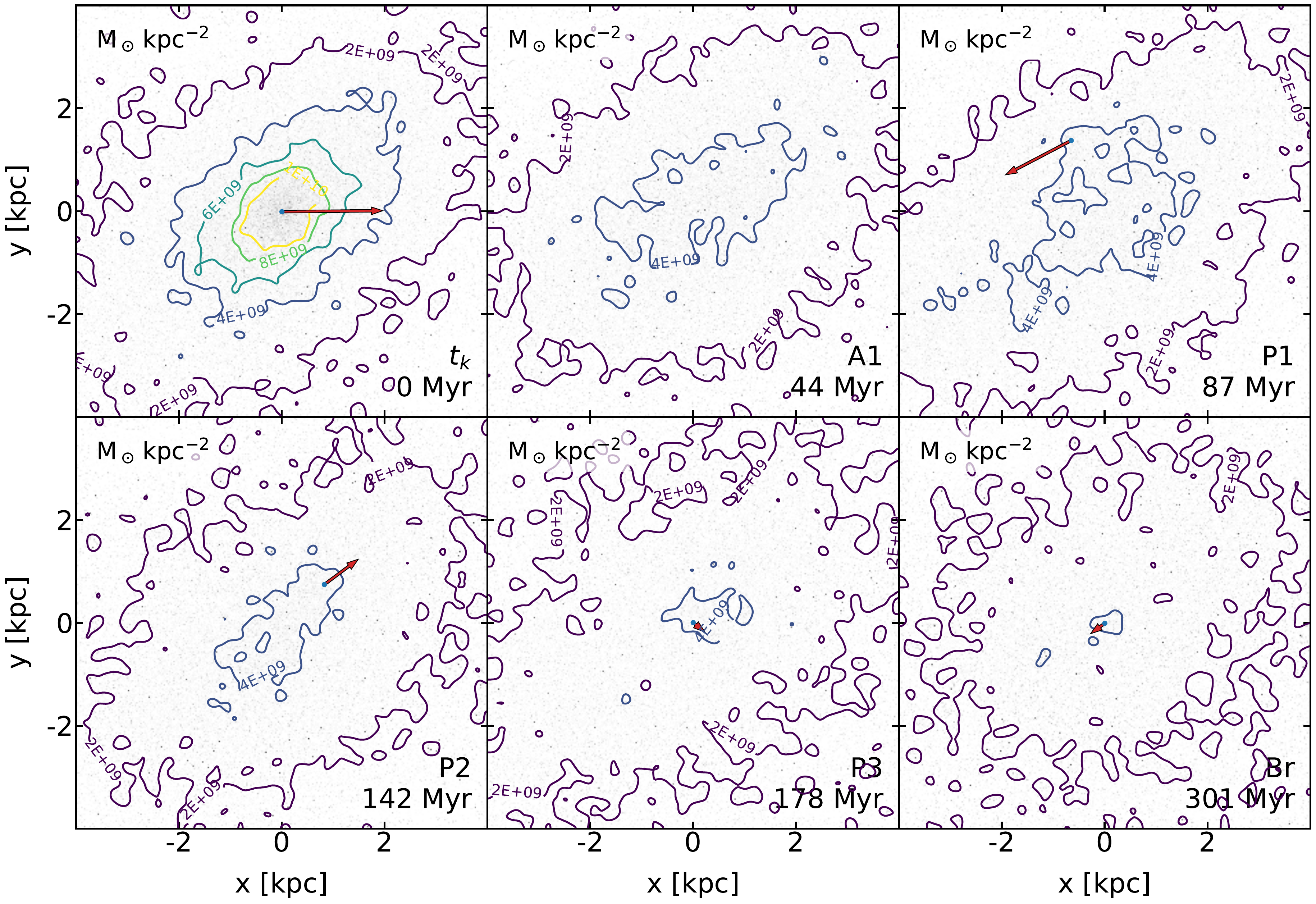}
     \caption{Contour plots of NGC 6166 with $\vk=0.5 \,\ve$. The blue dot is the position of the SMBH remnant, which leaves the area of the plot towards apocenter. The red arrow indicates the direction and magnitude of its velocity. An area of high density is present around the SMBH after the kick, and persists at ``Br", surrounded by a flat core.}
    \label{fig:contour_NGC6166_0.5}
\end{figure*}

The combination of the trajectories and density profiles suggests the SMBHs may be bringing the stars to the nucleus. This is supported by density contour plots, such as for the nucleus of NGC 6166 after a GW kick with $\vk/\ve$ of $0.5$ (Figure \ref{fig:contour_NGC6166_0.5}). After the kick, the flattening of the core as the SMBH moves away from the COM of the galaxy is evident, and the density continues to decrease, and the core to enlarge, with successive passages of the SMBH. There is also a higher density region around the SMBH, which moves out with it, and then returns to settle back at the center with the SMBH, surrounded by a flat core. This is likely due to the combination of the stars being dragged back to the dynamical center, and heating by dynamical friction.

\begin{figure*}
    \centering
    \includegraphics[width = 0.85\textwidth]{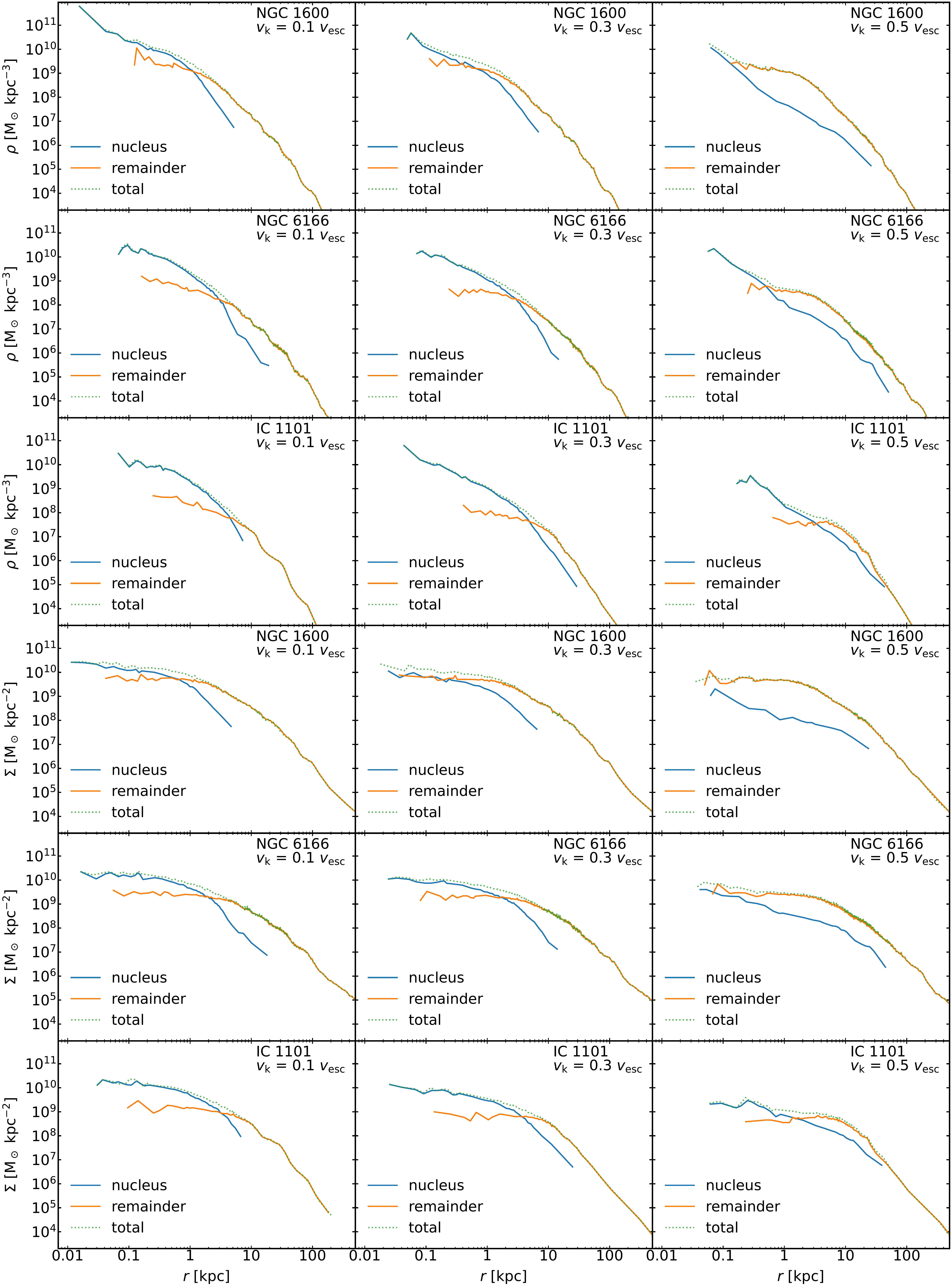}
    \caption{Volume (top three rows) and surface density (bottom three rows) profiles when the SMBH remnant has settled to Brownian motion for NGC 1600, NGC 6166 and IC 1101. Columns from left to right are for kicks with $\vk/\ve$ of $0.1$, $0.3$ and $0.5$ respectively. Profiles labeled ``nucleus" indicate those for particles that were within the influence radius of the SMBH at 1st apocenter (blue), and all other particles as ``remainder" (orange). The combined profiles are labeled ``total" (green dotted). The prominent increases in central density in NGC 1600 with $\vk=0.3 \,\ve$, and for NGC 6166 and IC 1101 with $\vk=0.5 \,\ve$ are seen to be due to the ``nucleus" particles.}
    \label{fig:nuc_select}
\end{figure*}

To confirm that the central density increase is due to this ``BH dragging" mechanism, we tagged
the stars within the SMBH sphere of influence at first apocenter and then plotted their density profiles at the end of the kick, when the SMBH reached Brownian motion, separately from the remaining stars (Figure \ref{fig:nuc_select}). These profiles, calculated using spherical shells, show three distinct scenarios: First, in the cases where the kick is relatively small and no increased central density is seen overall (eg. the $\vk/\ve$ of $0.1$ kicks and the $\vk/\ve$ of $0.3$ kicks for NGC 6166 and IC 1101), the tagged stars combine with the remainder to form a smooth profile; Second, for some of the largest kicks with no increased central density (eg. the $\vk/\ve$ of $0.5$ kicks for A2147-BCG and A2261-BCG), the tagged stars have a much lower density than the remainder; Finally, the most interesting cases are between these extremes (eg. the $\vk/\ve=0.3$ kick for NGC 1600 and the $\vk/\ve=0.5$ kicks for NGC 6166 and IC 1101), where the tagged stars rise steeply centrally above the flattened profile of the remainder, to form the increased nuclear density. 

\begin{figure}
    \includegraphics[width = \columnwidth]{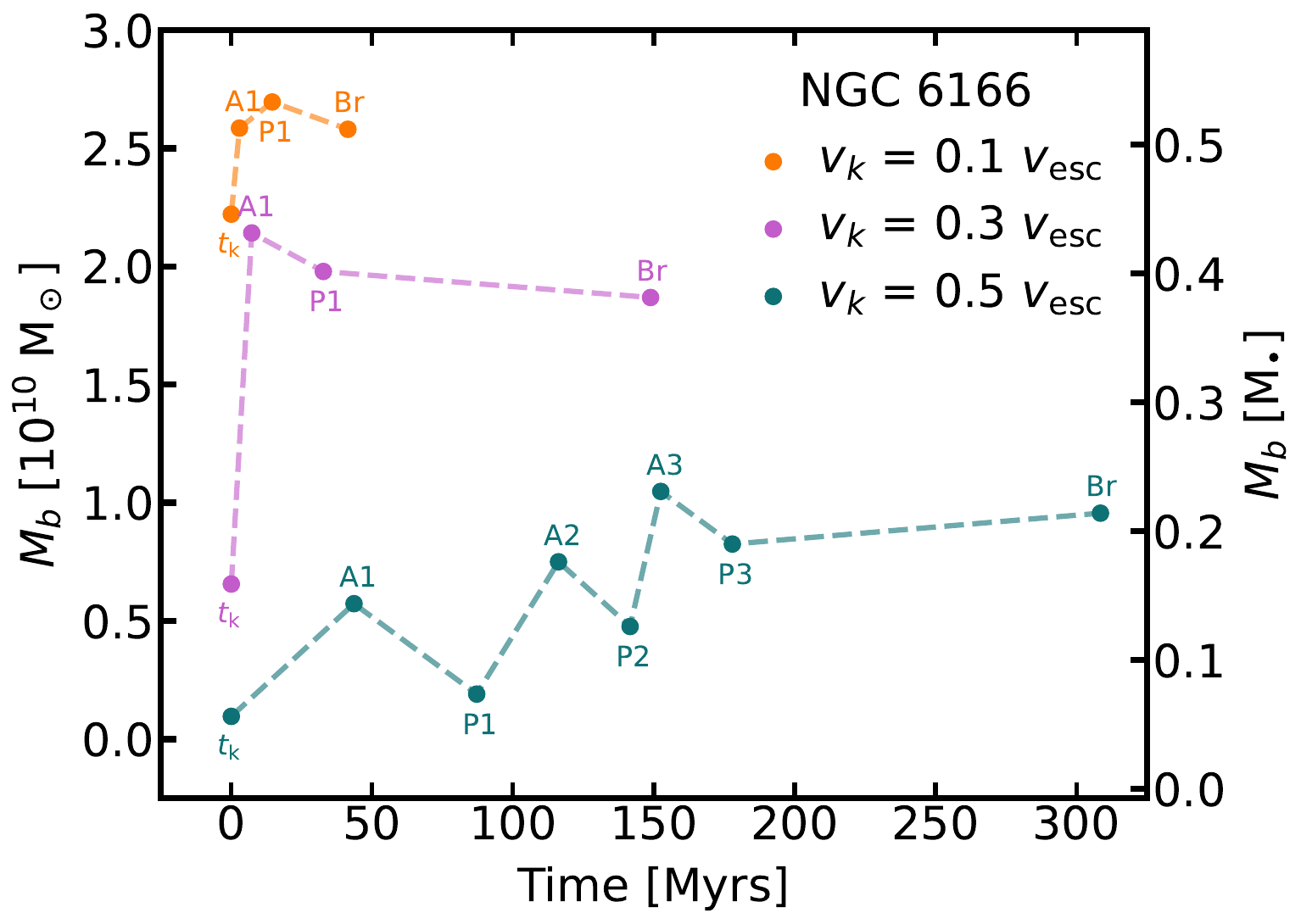}
     \caption{Bound mass to the SMBH remnant (here ``A" and ``P" indicate  apocenter and pericenter passages, followed by the number of passages; ``Br" indicates the SMBH remnant has settled into Brownian motion) for galaxy NGC 6166 with $\vk/\ve=0.1, 0.3, 0.5$. The bound mass is lower at higher $\vk/\ve$, for comparable phases after the kick. For the slower kicks, the bound mass quickly rises as the SMBH slows and remains high as the motion of the SMBH remnant is rapidly damped. For the $\vk/\ve=0.5$ kick, the bound mass is lower at each pericenter than the corresponding apocenter, as the SMBH speeds up and slows down.}
    \label{fig:bound_mass}
\end{figure}

The mass of stars bound to the SMBH remnant was calculated as the total mass of stars with negative total energy, that are also located within the influence radius of the SMBH ($\ri$). The latter is defined as the radius enclosing stars with twice the mass of the SMBH. Figure \ref{fig:bound_mass} shows the bound mass ($\mb$) for NGC 6166 with $\vk$ from $0.1$ to $0.5$ of $\ve$ at the time of the kick and at each apocenter and pericenter following it. The higher the GW kick velocity, the lower the bound mass at corresponding time-points. For the kicks with $\vk/\ve$ of $0.1$ and $0.3$, the bound mass increases quickly as the motion of the SMBH remnant is damped and remains high. However, for the $\vk = 0.5 \, \ve$ kick, the bound mass increases as the SMBH slows to each apocenter, and then drops as the remnant increases speed towards each pericenter.

\begin{deluxetable*}{lcccccccc}
\tablecaption{MCMC Core-S\'ersic fits.}
\label{tab:mcmc_kicks}    
\tablehead{
\colhead{Galaxy} & \colhead{$\vk/\ve$} & \colhead{$\vk$} & \colhead{$\rb$} & \colhead{$\gamma$} & \colhead{$\alpha$} & \colhead{n} & \colhead{$\re$} & \colhead{log $(\Sigma_b/10^9 $M$_\odot $kpc$^{-2})$} \\
            & & \colhead{(km s$^{-1}$)} & \colhead{(kpc)} & & & & \colhead{(kpc)} &}
\startdata
NGC 1600    & $0.0$                           & 0         & $0.68\,^{+0.04}_{-0.05}$    & $0.30\,^{+0.04}_{-0.04}$    & $3.0\,^{+0.5}_{-0.4}$       & $6.0\,^{+0.1}_{-0.2}$       & $22.3\,^{+0.4}_{-0.7}$  & $23.07\,^{+0.05}_{-0.05}$           \\& $0.1$                         & 309              & $0.84\,^{+0.03}_{-0.04}$  & $0.34\,^{+0.03}_{-0.03}$  & $4.7\,^{+0.7}_{-0.5}$     & $6.29\,^{+0.01}_{-0.01}$     & $22.35\,^{+0.09}_{-0.04}$    & $22.81\,^{+0.04}_{-0.03}$     \\ 
            & $0.3$                         & 929               & $1.17\,^{+0.14}_{-0.11}$  & $0.22\,^{+0.06}_{-0.12}$  & $3.37\,^{+2.5}_{-1.1}$   & $7.67\,^{+2.57}_{-1.52}$  & $26.40\,^{+5.86}_{-5.06}$ & $22.37\,^{+0.02}_{-0.02}$     \\
            & $0.5$                         & 1548              & $1.74\,^{+0.14}_{-0.13}$  & $0.09\,^{+0.07}_{-0.06}$  & $4.8\,^{+2.5}_{-1.4}$     & $7.6\,^{+2.4}_{-1.9}$     & $22.4\,^{+0.1}_{-0.1}$    & $21.92\,^{+0.02}_{-0.02}$     \\     
\noalign{\smallskip}
A2147-BCG   & $0.0$                        & 0         & $1.32\,^{+0.15}_{-0.16}$    & $0.23\,^{+0.06}_{-0.07}$    & $1.4\,^{+0.2}_{-0.1}$       & $6.3\,^{+0.1}_{-0.1}$       & $24.7\,^{+2.5}_{-1.4}$  & $21.95\,^{+0.10}_{-0.09}$           \\ & $0.1$                        & 323              & $2.29\,^{+0.03}_{-0.04}$  & $0.52\,^{+0.03}_{-0.03}$  & $2.8\,^{+0.3}_{-0.3}$     & $6.33\,^{+0.05}_{-0.08}$     & $23.3\,^{+0.8}_{-0.4}$    & $21.27\,^{+0.05}_{-0.05}$     \\ 
            & $0.3$                         & 971              & $2.74\,^{+0.29}_{-0.26}$    & $0.23\,^{+0.05}_{-0.09}$    & $3.3\,^{+1.8}_{-0.9}$ & $7.8\,^{+3.2}_{-2.1}$       & $23.1\,^{+5.7}_{-3.0}$    & $21.07\,^{+0.02}_{-0.02}$   \\
            & $0.5$                         & 1619              & $3.65\,^{+0.01}_{-0.10}$  & $0.20\,^{+0.03}_{-0.03}$  & $4.7\,^{+0.6}_{-0.5}$ & $6.3\,^{+0.1}_{-0.1}$     & $24.6\,^{+2.5}_{-1.4}$    & $20.47\,^{+0.02}_{-0.02}$     \\
\noalign{\smallskip}
NGC 6166   & $0.0$                         & 0                 & $0.90\,^{+0.09}_{-0.09}$    & $0.20\,^{+0.05}_{-0.06}$    & $1.15\,^{+0.05}_{-0.05}$    & $9.0\,^{+0.1}_{-0.1}$       & $60.1\,^{+0.2}_{-0.1}$  & $22.79\,^{+0.08}_{-0.08}$           \\
            & $0.1$                         & 331              & $1.12\,^{+0.05}_{-0.05}$  & $0.48\,^{+0.03}_{-0.03}$  & $4.8\,^{+0.7}_{-0.6}$     & $9.08\,^{+0.01}_{-0.03}$     & $60.3\,^{+0.6}_{-0.3}$    & $22.42\,^{+0.04}_{-0.04}$     \\ 
            & $0.3$                         & 994              & $1.83\,^{+0.04}_{-0.04}$  & $0.49\,^{+0.02}_{-0.02}$  & $8.9\,^{+1.9}_{-1.2}$ & $9.09\,^{+0.01}_{-0.01}$     & $60.1\,^{+0.2}_{-0.1}$    & $21.74\,^{+0.02}_{-0.02}$ \\
            & $0.5$                         & 1657              & $3.35\,^{+0.21}_{-0.23}$  & $0.21\,^{+0.06}_{-0.10}$   & $6.9\,^{+1.9}_{-1.7}$ & $8.23\,^{+3.96}_{-3.09}$     & $48.1\,^{+18.3}_{-4.1}$    & $21.09\,^{+0.01}_{-0.01}$ \\            
\noalign{\smallskip}
4C +74.13   & $0.0$                         & 0                 & $1.22\,^{+0.06}_{-0.06}$    & $0.24\,^{+0.03}_{-0.04}$    & $5.0\,^{+0.9}_{-0.6}$    & $3.89\,^{+0.01}_{-0.02}$    & $16.9\,^{+0.1}_{-0.1}$  & $22.40\,^{+0.03}_{-0.03}$           \\
            & $0.1$                         & 339               & $1.22\,^{+0.06}_{-0.06}$  & $0.24\,^{+0.03}_{-0.04}$  & $5.0\,^{+0.9}_{-0.6}$ & $3.89\,^{+0.01}_{-0.02}$    &$16.9\,^{+0.1}_{-0.1}$     & $22.40\,^{+0.03}_{-0.03}$     \\
            & $0.3$                         & 1019              & $2.53\,^{+0.06}_{-0.06}$  & $0.28\,^{+0.02}_{-0.02}$  & $6.3\,^{+1.0}_{-0.7}$ & $3.87\,^{+0.02}_{-0.05}$     & $17.0\,^{+0.2}_{-0.1}$    & $21.62\,^{+0.02}_{-0.02}$ \\
            & $0.5$                         & 1698              & $3.54\,^{+0.08}_{-0.08}$  & $0.17\,^{+0.02}_{-0.02}$  & $5.2\,^{+0.6}_{-0.5}$ & $3.83\,^{+0.05}_{-0.08}$     & $17.2\,^{+0.6}_{-0.3}$    & $21.10\,^{+0.02}_{-0.02}$ \\
\noalign{\smallskip}
A2261-BCG  & $0.0$                          & 0             &$1.32\,^{+0.04}_{-0.05}$    & $0.24\,^{+0.02}_{-0.02}$    & $62.4\,^{+25.8}_{-28.7}$    & $2.15\,^{+0.02}_{-0.02}$    & $13.1\,^{+0.1}_{-0.1}$  & $22.98\,^{+0.02}_{-0.02}$           \\
            & $0.1$                       & 356              & $0.96\,^{+0.05}_{-0.05}$  & $0.22\,^{+0.02}_{-0.03}$  & $52.9\,^{+32.0}_{-31.6}$ & $2.199\,^{+0.001}_{-0.001}$     & $12.905\,^{+0.008}_{-0.004}$    & $22.79\,^{+0.03}_{-0.03}$     \\
            & $0.3$                       & 1070              & $2.08\,^{+0.05}_{-0.06}$  & $0.25\,^{+0.02}_{-0.02}$  & $32.9\,^{+40.9}_{-17.2}$ & $2.08\,^{+0.05}_{-0.06}$  & $12.91\,^{+0.02}_{-0.01}$    & $22.20\,^{+0.02}_{-0.02}$     \\
            &$0.5$                         & 1784              & $2.65\,^{+0.06}_{-0.07}$  & $0.15\,^{+0.02}_{-0.02}$  & $8.8\,^{+2.0}_{-1.3}$ & $2.19\,^{+0.01}_{-0.01}$  & $12.93\,^{+0.05}_{-0.03}$    & $21.90\,^{+0.02}_{-0.02}$     \\ 
\noalign{\smallskip}
IC 1101     & $0.0$                       & 0                & $0.61\,^{+0.07}_{-0.07}$     & $0.23\,^{+0.09}_{-0.11}$  & $3.1\,^{+1.0}_{-0.7}$ & $5.2\,^{+0.2}_{-0.1}$  & $11.7\,^{+0.2}_{-0.3}$  & $22.55\,^{+0.09}_{-0.08}$           \\
            & $0.1$                       & 318              & $0.63\,^{+0.12}_{-0.10}$     & $0.12\,^{+0.09}_{-0.07}$  & $1.1\,^{+0.1}_{-0.1}$ & $5.4\,^{+0.4}_{-0.3}$  & $11.6\,^{+0.3}_{-0.4}$  & $22.46\,^{+0.11}_{-0.14}$           \\
            & $0.3$                       & 956              & $4.78\,^{+0.36}_{-0.34}$     & $0.52\,^{+0.05}_{-0.04}$  & $1.2\,^{+0.2}_{-0.1}$ & $5.5\,^{+0.3}_{-0.4}$  & $11.5\,^{+0.3}_{-0.4}$  & $20.00\,^{+0.09}_{-0.09}$           \\
            & $0.5$                       & 1594             & $6.81\,^{+0.38}_{-0.40}$     & $0.21\,^{+0.06}_{-0.10}$  & $6.4\,^{+2.1}_{-1.6}$ & $7.7\,^{+2.3}_{-2.3}$  & $16.1\,^{+1.3}_{-1.3}$  & $19.12\,^{+0.03}_{-0.03}$           \\
\enddata
\tablecomments{The fits were performed on surface density profiles at the time the SMBH reaches Brownian motion.}
\end{deluxetable*}

\begin{figure}
    \includegraphics[width = \columnwidth]{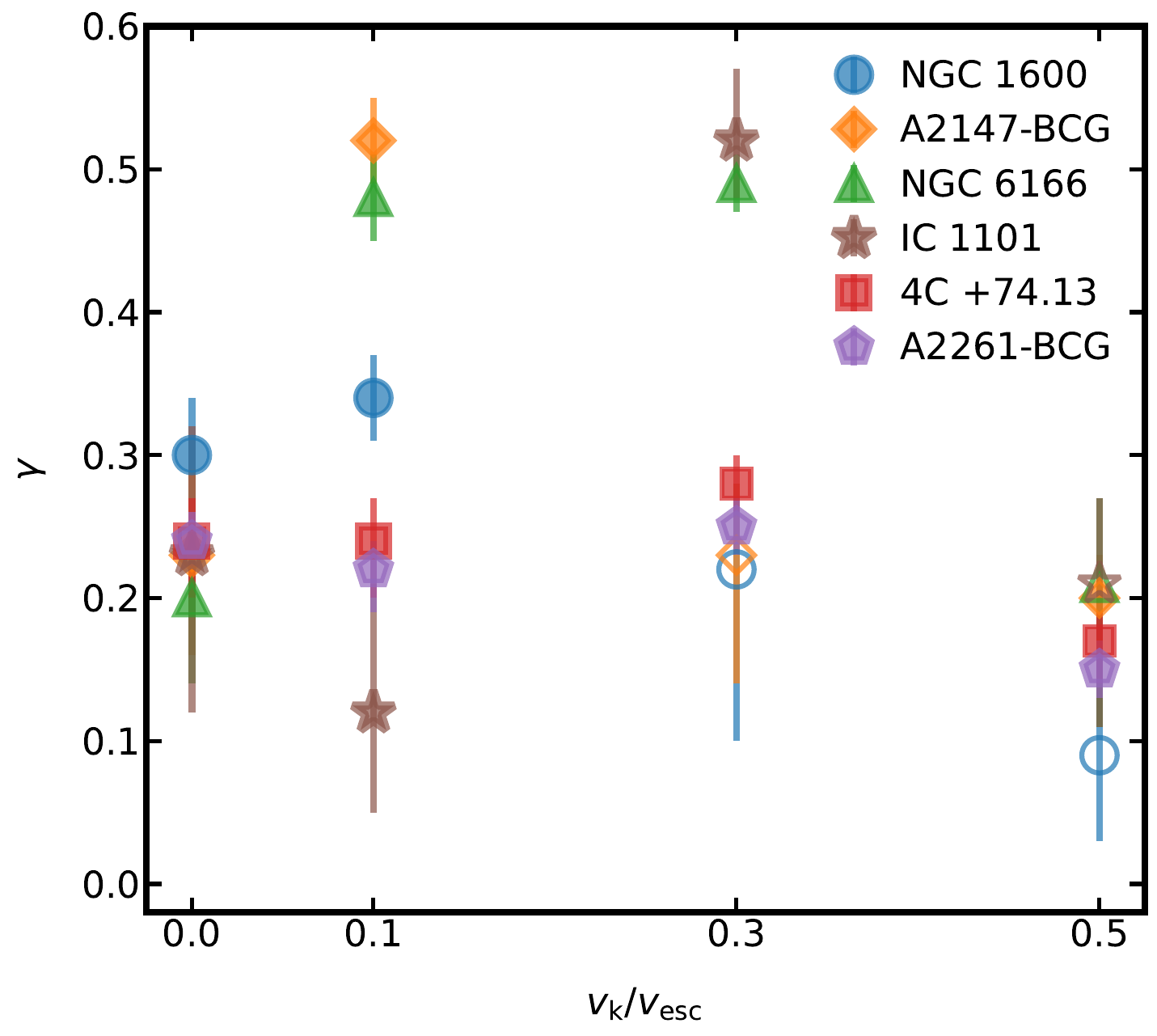}
     \caption{Inner logarithmic slope ($\gamma$) as a function of the ratio of kick velocity ($\vk$) to escape speed ($\ve$) for all six galaxy remnants. A $\vk/\ve$ of zero indicates scouring only. If the NSC cannot be excluded from the fit at $\vk/\ve$ of $0.1$ or $0.3$, there is a spike in $\gamma$ to $\sim 0.5$ due to its presence. Where the NSC is excluded (unfilled markers), the relation between $\gamma$ and $\ve$ is restored.}
    \label{fig:gamma_vk}
\end{figure}

\begin{figure*}
    \centering
    \includegraphics[width = \textwidth]{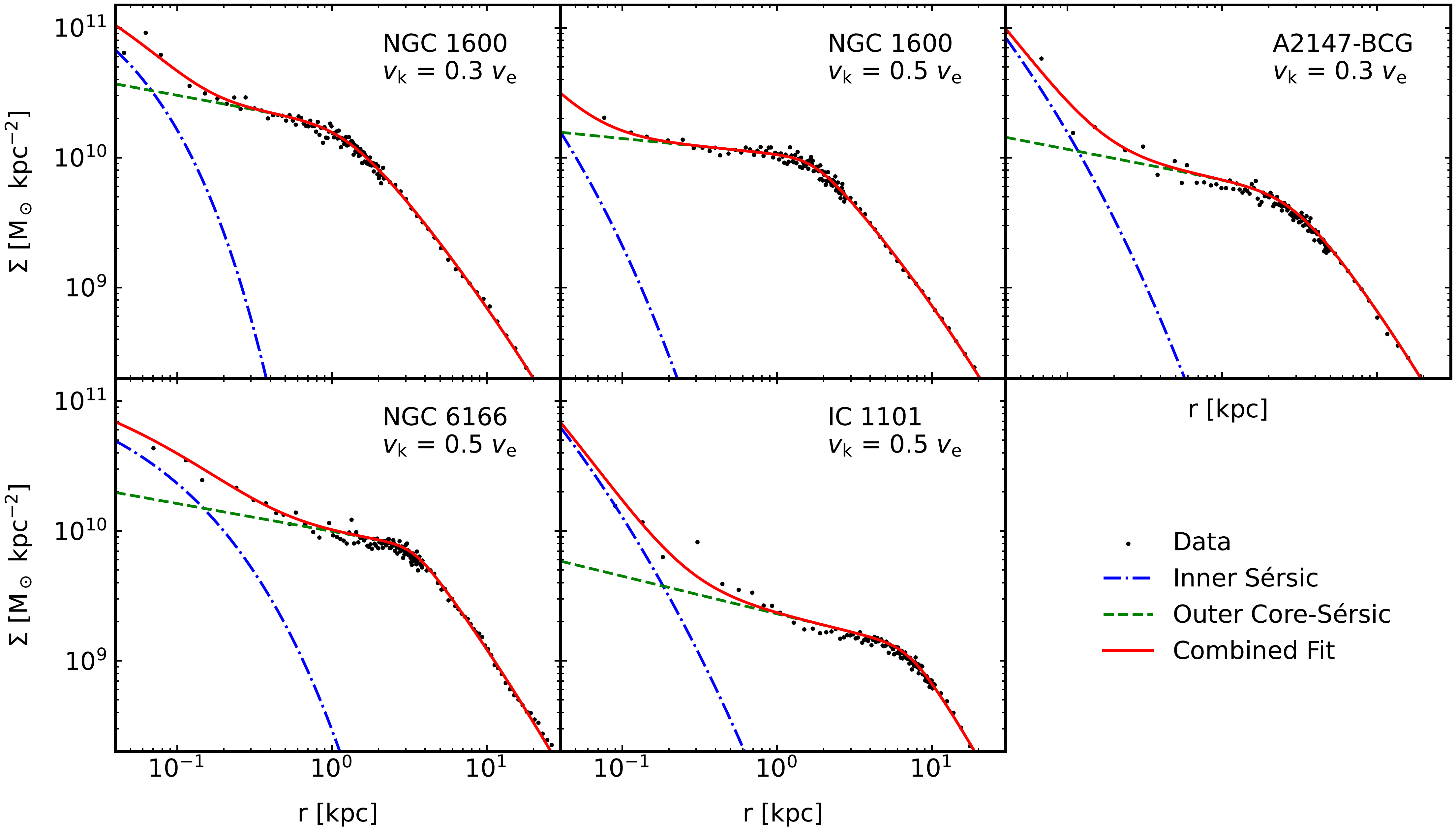}
    \caption{Surface density profiles and combined inner \sersic and outer \cs fits for the five most prominent NSCs: NGC 1600 after kicks with $\vk=0.3$ and $0.5$ of $\ve$ (top left and top center, respectively), A2147 with $\vk=0.3 \,\ve$ (top right), and NGC 6166 and IC 1101 with $\vk=0.5 \,\ve$ (bottom left and bottom center, respectively). The binned data are shown as black points, the inner \sersic component of the fit as a blue dot-dashed line, the outer \cs component as a green dashed line and the combined fit as a solid red line. The effective radius for the \sersic fit is used as the radius of the NSC.}
    \label{fig:double_fits}
\end{figure*}

Markov chain Monte Carlo (MCMC) fitting of the surface density profiles was performed as in \citetalias{khonji2024core} using the \cs profile \citep{graham2003new}. The results of the fits, given in Table \ref{tab:mcmc_kicks}, confirm a progressive increase in $\rb$ with increasing $\vk/\ve$. As shown in \citetalias{khonji2024core}, NGC 1600 and A2147-BCG do not require GW recoil to achieve their observed sizes, but NGC 6166 and A2261-BCG require a kick with $\vk/\ve \lesssim 0.5$. For the two remaining galaxies, a GW kick with $\vk/\ve$ between $0.1$ and $0.3$ is required for both  4C +74.13 and IC 1101. A key finding in \citetalias{khonji2024core} was that GW kicks resulted in uniquely flat cores, especially in 3D. With the lower speed kicks in this study, we find that $\gamma$ may actually increase (see Figure \ref{fig:gamma_vk}). This occurs when the \mbox{nuclear} density blends smoothly into the overall profile, and is most prominent for $\vk/\ve=0.1$ in A2147-BCG and NGC 6166, and for $\vk/\ve=0.3$ in NGC 6166 and IC 1101. In these cases, $\gamma\sim 0.5$, indicating the formation of a central cusp. For the five most prominent NSCs, there are central cusps present but, due to the surrounding flat cores, they could be excluded from the fit so that the flat signature for high kicks is retained. 

\subsection{Properties of the NSCs}

Here, we analyze the clusters of stars bound to the MBH, but we return to the question whether these constitute NSCs in the observational sense in Section \ref{sec:discussion}. 

The properties of the NSCs are shown in Table \ref{tab:nsc_prop}. Observers generally fit the light profile of such systems with the combination of an inner and outer profile, and use the effective radius of the inner profile as the radius of the NSC \citep [e.g.][] {dullo2019most}. To obtain a comparable radius for the simulation NSCs, we fitted the surface density profile of the remnant, once the SMBH had reached Brownian motion, using an inner \sersic profile and an outer \cs profile (Figure \ref{fig:double_fits}). The effective radius of the \sersic fit was taken as the radius of the NSC ($\rn$). There is good agreement, within a factor of $2$\mbox{--}$3$, between $\rn$ and the characteristic length $\rk$, the theoretical linear extent of a bound cluster to an ejected SMBH \citep{merritt2009hypercompact}, given by
\begin{equation}
\label{eq:rk}
\rk = \frac{G\mbh}{\vk^2} \,.
\end{equation} 

\begin{deluxetable*}{lcclcccccccc}
\tablecaption{NSC properties}
\label{tab:nsc_prop}    
\tablehead{
\colhead{Galaxy} & \colhead{$\vk/\ve$}    & \colhead{$\rk$} &  \colhead{$\rn$}  & \colhead{$\mb(\tk)$}& \colhead{$\mn$} & \colhead{$\mn/\mbh$} & \colhead{$\sigma_N$} & \colhead{$N_N$} & \colhead{$\fhr$} & \colhead{$\en$}\\
&   & \colhead{$(\pc)$}  & \colhead{$(\pc)$} & \colhead{($10^{8} $M$_\odot$)} & \colhead{($10^{8} $M$_\odot$)} & & ($\kms$) & & &}
\startdata
NGC 1600    & 0.3   & \phantom{5}80 & \phantom{5}$80 \,^{+30}_{-20}$    & 15.0              & \phantom{5}$5.8 \,^{+2.2}_{-1.5}$    & $0.034 \,^{+0.013}_{-0.009}$& 436    & 315             & 0.95 & 0.107\\
            & 0.5   & \phantom{5}30 & \phantom{5}$40 \,^{+20}_{-20}$    & \phantom{5}1.4    & \phantom{5}$0.5 \,^{+0.3}_{-0.3}$    & $0.003\,^{+0.002}_{-0.002}$ & 405    & \phantom{5}31   & 0.90 & 0.238 \\
\noalign{\smallskip}
A2147-BCG   & 0.3   & 120           & $220 \,^{+120}_{-70}$             & 14.9              & $13.3 \,^{+7.2}_{-4.2}$              & $0.050\,^{+0.028}_{-0.016}$ & 411    & 349             & 0.93 & 0.058\\ 
\noalign{\smallskip}
NGC 6166    & 0.5   & \phantom{5}70 & $150 \,^{+80}_{-50}$              & \phantom{5}9.6    & \phantom{5}$8.2 \,^{+4.3}_{-2.7}$    & $0.017\,^{+0.009}_{-0.006}$ & 508             & 313             & 0.96  & 0.058\\
\noalign{\smallskip}
IC 1101     & 0.5   & 190           & $470 \,^{+170}_{-330}$            & 30.4              & $38.4 \,^{+13.9}_{-27.0}$             & $0.035\,^{+0.013}_{-0.025}$ & 637              & 961             & 0.84 & 0.076\\
\enddata
\tablecomments{NSC properties: $\rk$ is the theoretical linear extent of a bound cluster to an ejected SMBH; $\rn$ is the effective radius of the inner S\'ersic fit and $\mb$ is the bound mass of the SMBH; $\mn$ and $N_N$ are the mass and particle number within $\rn$, respectively; $\sigma_N$ is the mean stellar velocity dispersion within 0.595 kpc at $\tbr$; $\fhr$ is the fraction of $N_N$ that are of the two highest resolutions; $\epsilon$ is the ellipticity of the stars within $\rn$.}
\end{deluxetable*}

\begin{figure}
    \includegraphics[width = \columnwidth]{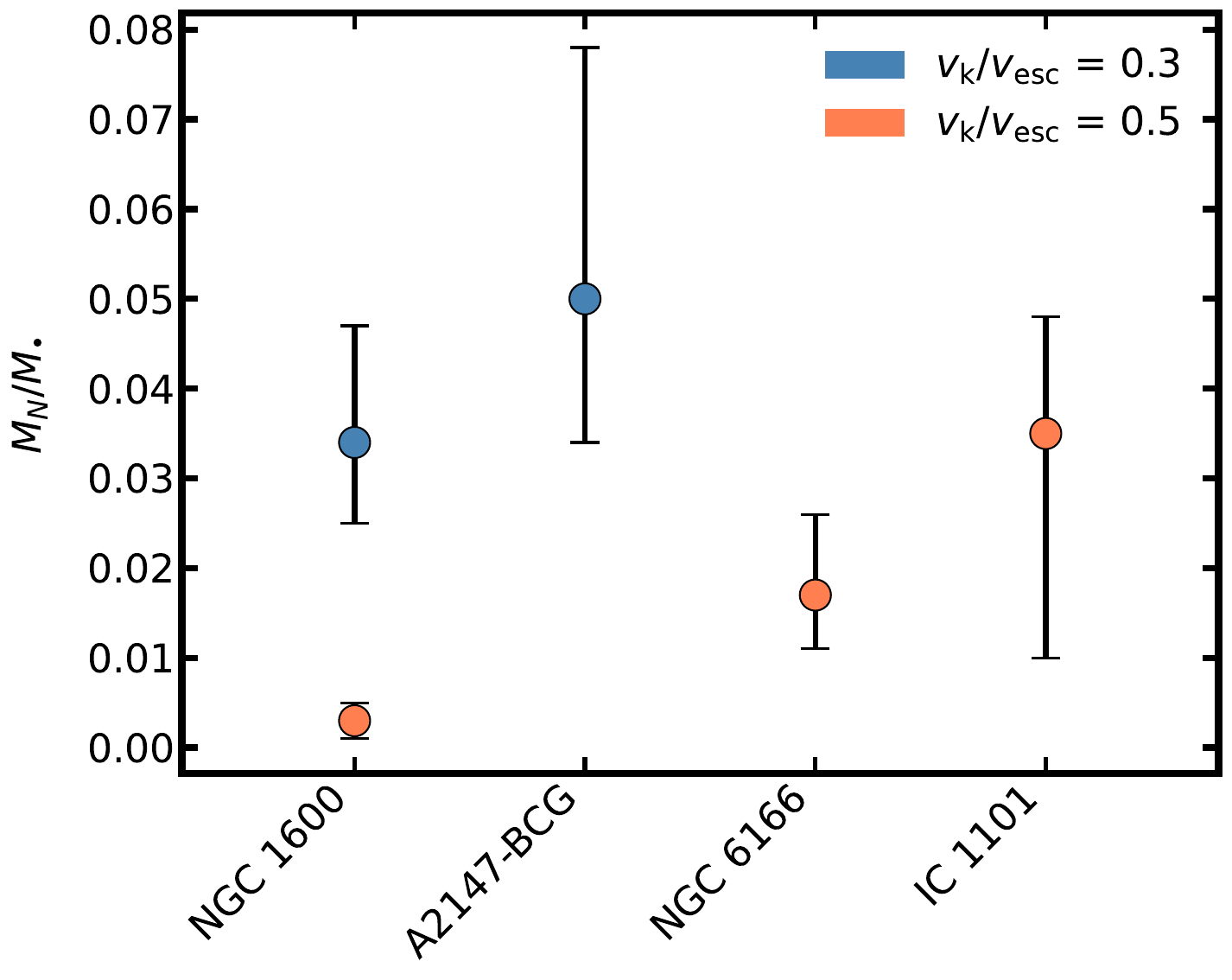}
     \caption{NSC mass ($\mn$) as a proportion of SMBH mass ($\mbh$) for the five most prominent NSCs. The galaxies are listed in order of increasing SMBH mass from left to right. Those forming after kicks with $\vk/\ve$ of $0.3$ and $0.5$ are shown in blue and orange, respectively. $\mn$ appears to scale faster than $\mbh$ and $\mn/\mbh$ may be smaller for kicks with $\vk/\ve$ of $0.5$ than $0.3$.}
    \label{fig:MN_bh_vk}
\end{figure}

\begin{figure}
    \includegraphics[width = \columnwidth]{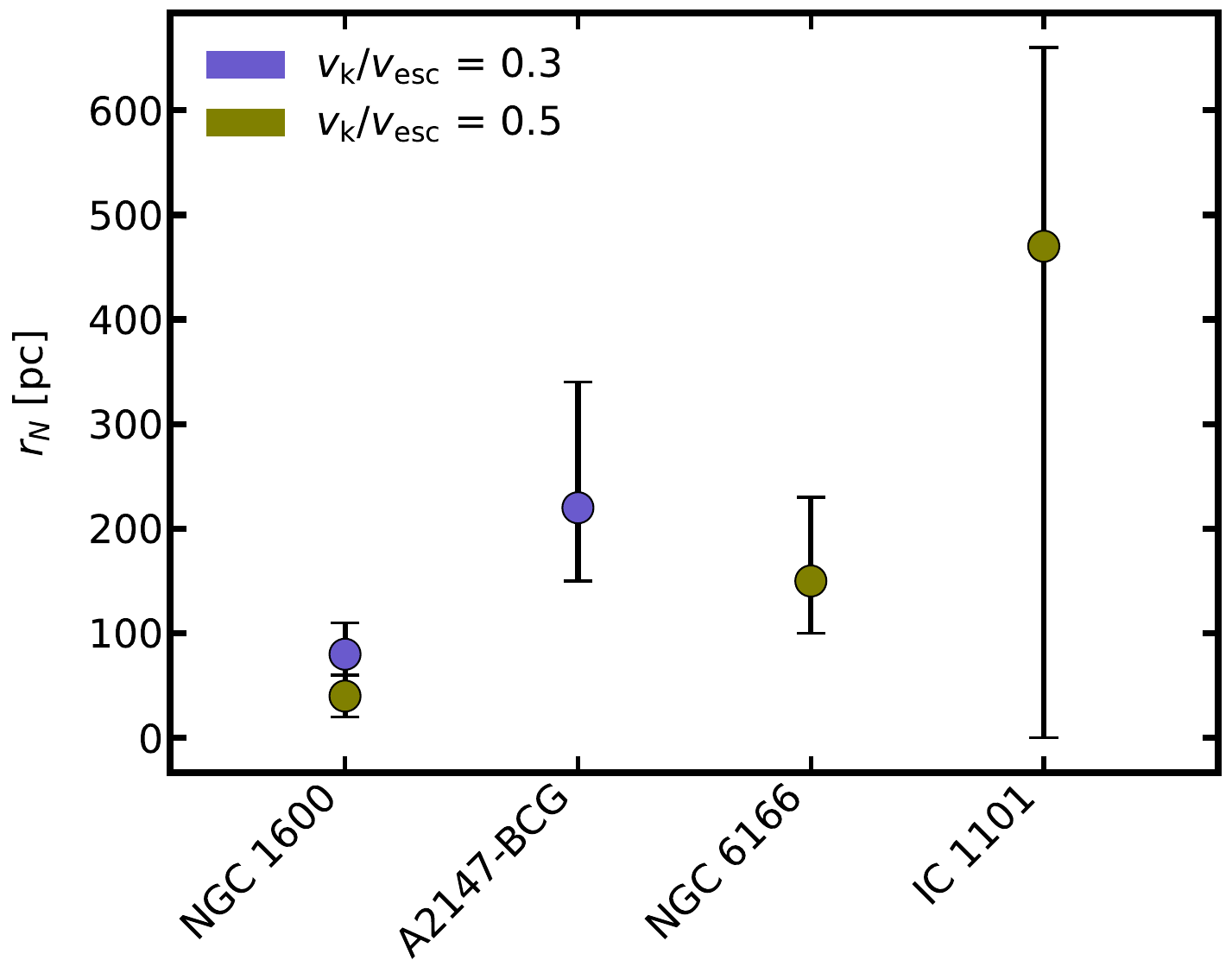}
     \caption{NSC radius ($\rn$) for the five most prominent NSCs. Those forming after kicks with $\vk/\ve$ of $0.3$ and $0.5$ are shown in purple and olive, respectively. The galaxies have increasing SMBH mass ($\mbh$) from left to right, and there appears to be a trend for increasing $\rn$ with $\mbh$.}    
     \label{fig:radius}
\end{figure}

\begin{figure*}
\centering

\begin{minipage}{0.49\textwidth}
  \includegraphics[width=\linewidth]{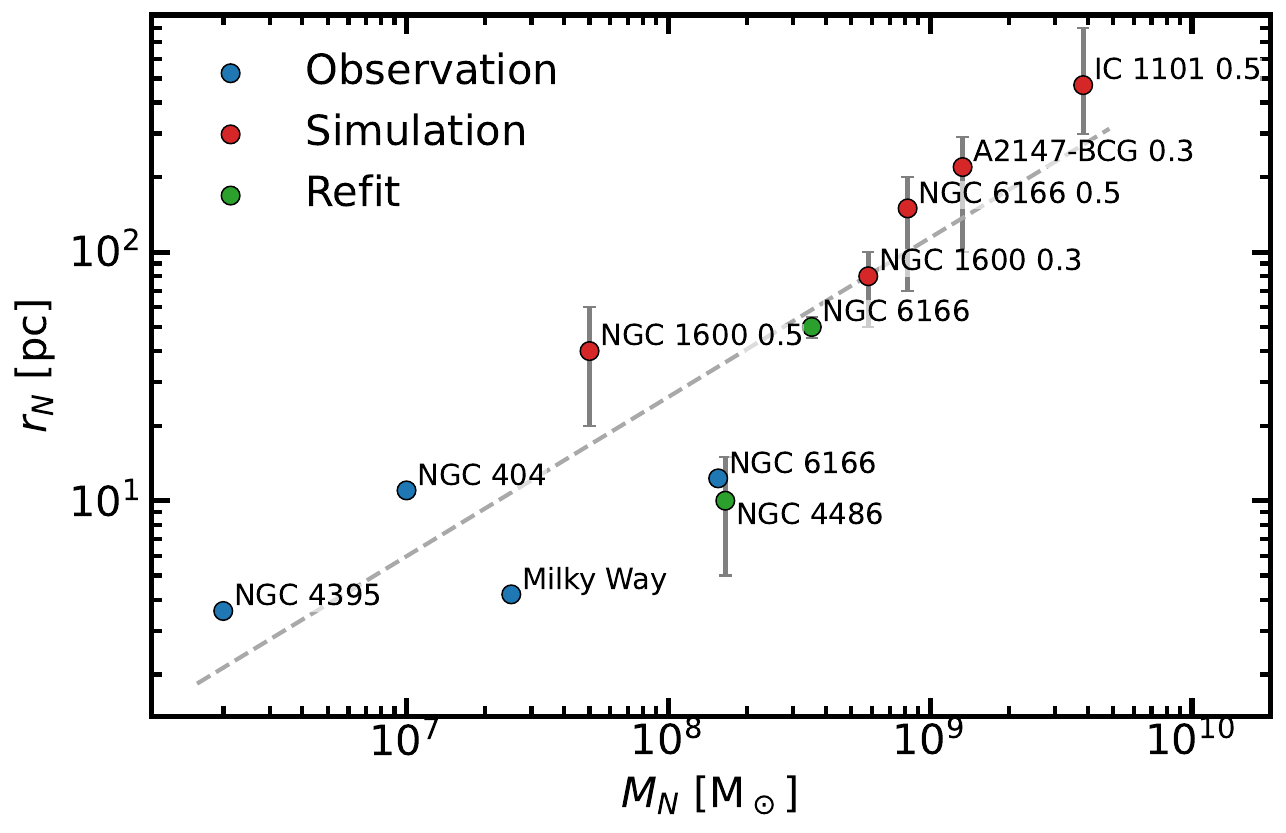}
\end{minipage}
\hfill
\begin{minipage}{0.49\textwidth}
  \includegraphics[width=\linewidth]{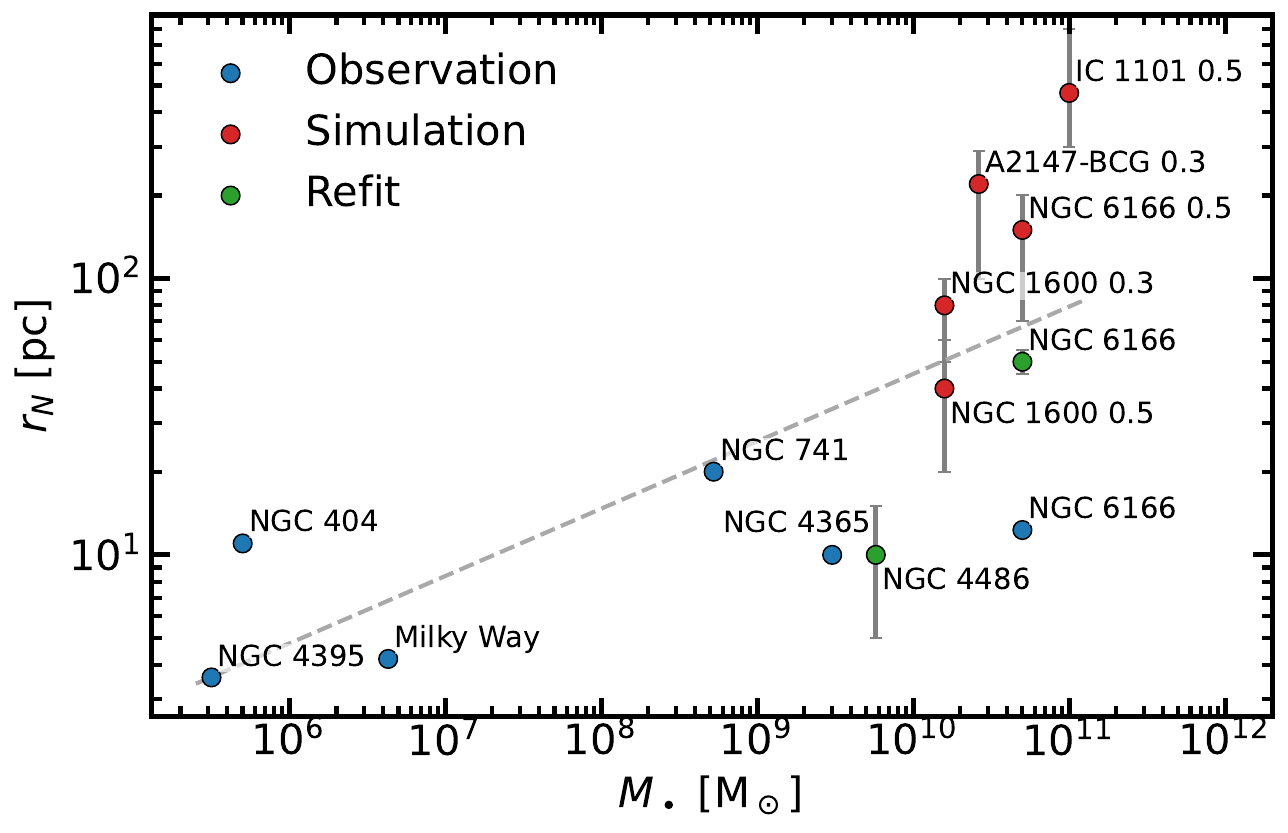}
\end{minipage}

\vspace{2mm}

\begin{minipage}{0.49\textwidth}
  \includegraphics[width=\linewidth]{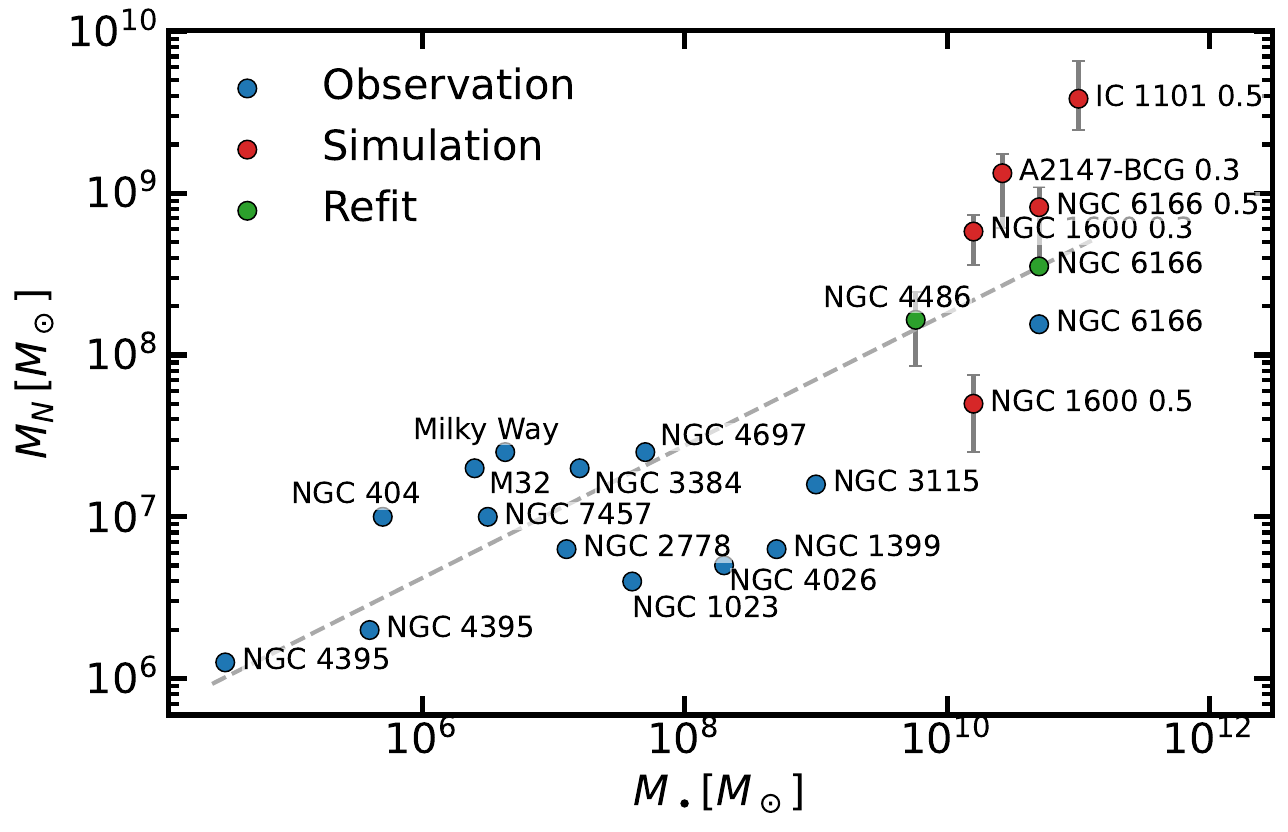}
\end{minipage}
\hfill
\begin{minipage}{0.49\textwidth}
  \includegraphics[width=\linewidth]{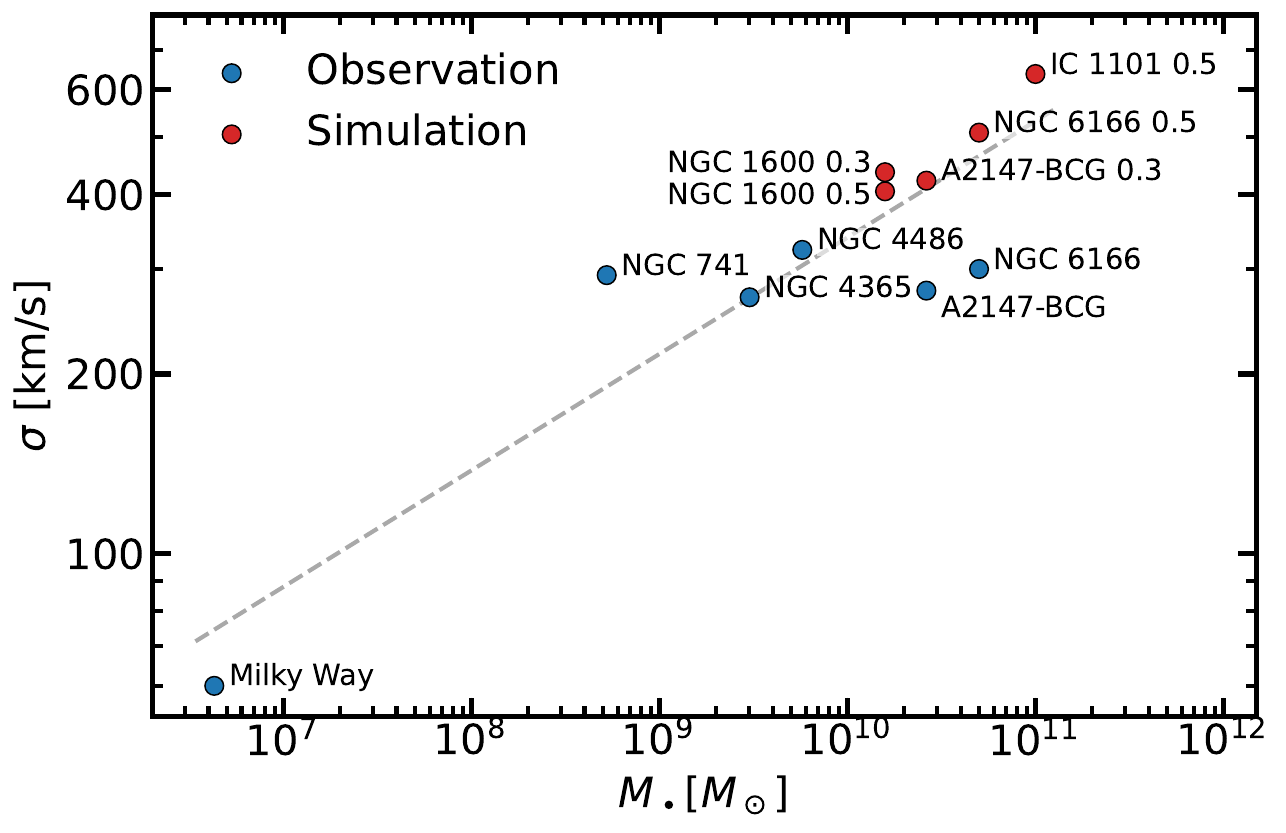}
\end{minipage}

\caption{NSC relations: NSC mass–radius relation (top left); SMBH mass relation with NSC radius (top right), NSC mass (bottom left), and velocity dispersion (bottom right). Blue points are observations from \cite{dullo2019most, schodel2014nuclear, feldmeier2014large, graham2009quantifying, seth2010ngc, den2015measuring}, red points are simulations from this study, and green points are refitted observations from \cite{dullo2019most}.}
\label{fig:relations}
\end{figure*}

Once $\rn$ is determined, the mass of the NSC ($\mn$) can be found as the mass within $\rn$. There is similarly good agreement between the bound mass at the time of the recoil kick $\mb(\tk)$ and $\mn$. The velocity dispersion of the NSC ($\sn$), is determined using
\begin{equation}
\sigma_N  = 
\frac{\int_0^K \sigma_{\text{los}}(R) \, \Sigma(R) \, R \, dR}
     {\int_0^K \Sigma(R) \, R \, dR} \, ,
\end{equation}
with $K$ set to $0.595$ kpc, the aperture used for $\sigma$ in the HyperLEDA database \citep{makarov2014hyperleda}. Finally, the ellipticity of the NSC ($\en$) is that of the stars within $\rn$. We find that the NSCs are mostly spherical ($\en \sim 0.06\mbox{--}0.24$).

Although limited by small number statistics, ratios of $\mn/\mbh$ (Figure \ref{fig:MN_bh_vk}) may be higher for a $\vk$ of $0.3 \, \ve$ than for $0.5 \, \ve$. This is most clearly apparent for NGC 1600, which forms an NSC in both cases, but with an order of magnitude difference in $\mn/\mbh$. However, since $M_N$ for a given $\vk/\ve$ appears to scale faster than $\mbh$, the values of $\mn/\mbh$ for the $0.5 \, \vk/\ve$ kicks in NGC 6166 and IC 1101 also appear low relative to those for the $0.3 \, \vk/\ve$ kicks for NGC 1600 and A2147-BCG. Similar trends are seen for $\rn$ (Figure \ref{fig:radius}), which appears to scale with $\mbh$ and may also be lower for higher $\vk/\ve$. 

\subsection{Relaxation Time}

To ensure the above results are not affected by relaxation, we estimate the relaxation time using 
\begin{equation}
    t_{\mathrm{relax}} = 0.34 \frac{\sigma_N^3}{G^2 m \rho \, \mathrm{ln} \Lambda} 
\end{equation}
\citep{binney2011galactic}, where $\sigma_N$ is the mean stellar velocity dispersion in the nuclear cluster, $\rho$ is its density and $\mathrm{ln} \, \Lambda$ is the Coulomb logarithm, where
\begin{equation}
    \Lambda = \frac{b_{\mathrm{max}}v^2_{\mathrm{typ}}}{2Gm} \, .
\end{equation}
Here, $b_{\mathrm{max}}$ is the maximum impact parameter and $v_{\mathrm{typ}}$ is the typical velocity of the cluster stars. This gives $t_{\mathrm{relax}} \approx 15$ Gyr at a radius of $1 \kpc$, far greater than the maximum time for the SMBH to settle into Brownian motion of $0.6$ Gyr.

\section{Discussion and Conclusions}
\label{sec:discussion}

We have studied the effect of relatively small GW recoil kicks following equal mass mergers and binary scouring to form some of the largest elliptical galaxies. We unexpectedly find that, unlike the flat cores which are a signature of GW recoil heating by larger kicks in \citetalias{khonji2024core}, the higher bound mass to the SMBH  typical of smaller kicks results in stars being pulled along with the SMBH remnant, a process we call ``BH dragging." 

The total mass of stars with negative total energy, that are also located within the influence radius of the SMBH, is taken as the bound mass ($\mb$). This increases with decreasing kick speed ($\vk$) at the time of the kick ($\tk$), at the first apocenter and pericenter, and at the time the SMBH reaches Brownian motion ($\tbr$, Figure \ref{fig:bound_mass}). Furthermore, for each kick, $\mb$ increases between $\tk$ and first apocenter, and remains higher than at $\tk$, showing how the NSC builds mass.

The final result depends on both the kick velocity and the galactic environment. Although the SMBH remnants have the highest bound mass for the smallest kicks studied, with $\vk/\ve=0.1$, there may be little change in the profile, as the SMBH moves only a small distance ($\lesssim 1 \kpc$) and is quickly damped, so heating is minimal. For the kicks with a $\vk$ of 0.3 or 0.5 of $\ve$, there is usually a reduction in overall density further out due to the greater distance traveled by the SMBH. The combination of the stars being dragged back to the dynamical center and heating by dynamical friction results in the nuclear density being visible and distinct from the remainder of the stellar bulge. However, there is a possible intermediate scenario, where the heating is not as effective, but the BH dragging still occurs, which can result in a cuspy profile, with a higher $\gamma$ than before the kick. 

We have shown that the simulation profiles can be decomposed into the stars that were bound to the SMBH remnant at first apocenter and the remainder, and that it is the former which can cause the prominent central nuclear density. This is clear evidence that the BH dragging mechanism is the cause for the formation of NSCs in the simulations, where the decomposition is straightforward to implement. However, this is obviously not possible in the case of observed surface brightness profiles. 

Where the nucleus is distinct, we understand that observers would fit this separately, often using a S\'ersic profile. For the initial \cs fits (Table \ref{tab:mcmc_kicks}), where it was sensible and practicable, we simply excluded the central density when fitting the Core-S\'ersic profile, but for the five runs with the most prominent NSCs, we have followed the practice of observers, by fitting a combined profile using inner \sersic and outer \cs components. This allowed us to obtain NSC masses and radii for comparison with observed NSCs. However, neither method is possible in the case of the cuspy profiles and so it is likely these galaxies would be classed as cusp rather than core galaxies. The increased cuspiness also leads to the possibility that galaxies after a small GW recoil would be classified as cusp rather than core galaxies.

To compare the properties of our NSCs with those observed, their radii, masses and velocity dispersions are plotted with examples from the literature (Figure \ref{fig:relations}). In addition, we refitted the surface brightness data for NGC 6166 and NGC 4486 from \citep{dullo2019most}, using the same procedure as for our simulations, to give an even better comparison. These will be discussed in detail in forthcoming work (N. Khonji et al., 2025, in preparation).

The NSC mass-radius relation shows that the simulations produce NSCs which are of an appropriate mass for their size, in comparison to observed NSCs. Although they are larger and more massive than most of the latter, there is some overlap, especially when the refitted galaxies are included. This is also the case for the relations between SMBH mass and NSC mass or radius, although the relation appears steeper for the simulations in the latter in comparison to the observed NSCs. The velocity dispersions of the simulated NSCs are higher than those observed, but within a factor of $\sim 2$. Overall, we believe the slightly larger properties are reasonable, given the extremely high SMBH masses and stellar masses used in the simulations. We also find that BH dragging results in mostly spherical NSCs.

A broader question is whether NSCs even exist in larger elliptical galaxies. \citet{cote2006acs} found no nucleation in galaxies brighter than $\sim$ $M_B$ -20.5, and suggest that any nuclei are erased by subsequent mergers or core formation. However, \cite{lauer2005centers} find increased central densities in 29\% of elliptical galaxies and \cite{dullo2024lemmings} find nuclei in 10-20 \% of core galaxies but 76 \% of cusp galaxies. It is interesting to note that BH dragging occurs after the merger and core formation so these processes would not erase the nucleus unless a subsequent merger occurs, and that it results in a more cuspy profile. Furthermore, we calculate the relaxation time for these large galaxies to all be $>$ 10 Gyr, so that they will persist, although the massive SMBHs are likely to be fed by tidal disruption events. Detection of such clusters in giant elliptical galaxies could be used as a probe of low velocity recoil kicks. We will examine this in future work.

Metallicity acts as an important differentiator between NSC formation mechanisms. Nuclei formed by infall of GCs should have lower metallicities than the remainder of the galaxy, whereas they should be higher in those from in situ star formation. BH dragging in gas-poor ellipticals would be expected to have the same metallicity as the central stars. \cite{neumayer2020nuclear} find that nuclei in elliptical galaxies above 10$^9 M_\odot$ generally have higher metallicities than their hosts, especially in spectroscopic observations, such that they favor in-situ star formation over GC infall in such massive ellipticals. Although we agree the data suggests formation by GC infall is less likely, the metallicity of galaxies has a general tendency to rise towards their centers \citep{fahrion2021diversity}, so that NSCs formed by BH dragging would be expected to have higher metallicities than the remainder of the galaxy. Low ellipticity is another potentially testable prediction of the mechanism. 

It is also possible that improved multiwavelength observations may allow detection of more massive nuclei in the most massive galaxies. For example, \cite{dullo2024lemmings} finds hybrid nuclei with an NSC and AGN are three times more frequent than previously observed \citep[e.g.][]{seth2008coincidence} in galaxies with stellar masses of 10$^{10.6}$-10$^{11.8} $ M$_\odot$, and finds NSC masses of up to 10$^{9.8}$. Observational mass estimation is also not straightforward and different methods may be used, but they are predominantly photometric, by 1D or 2D decomposition of the surface brightness profile, using multiple components, and conversion of luminosities into mass using colors and stellar population mass-to-light ratios \citep[e.g.][]{georgiev2016masses, dullo2024lemmings}. 

Of course, our results must be viewed within the context of the limitations of the simulations. As in all simulations which approximate the stellar density distribution with higher than stellar mass particles, collisional effects are necessarily approximated and are not as accurately modeled as in star-by-star simulations. Additionally, our stellar resolution is relatively low, as the simulations were not designed to look for NSC formation, which was an unexpected finding. Clearly, systems of this mass cannot currently be modeled on a star-by-star basis, but further study with higher resolution is planned to confirm the BH dragging process and examine it in more detail. Resolution could be enhanced further by using adaptive softening, allowing less conservative choices of mass shells for the mass-refinement scheme. Furthermore, our simulations assume that these galaxies contain so little gas that hydrodynamic simulations are unnecessary, but it may be interesting to examine the effect of adding a small density of gas in further work. 

However, if BH dragging is confirmed, there are two main implications. First, there is no reason that this process would not scale to smaller elliptical galaxies and SMBHs. BH-dragging may have occurred in such galaxies after major mergers, and be responsible for some of the NSCs already observed. Second, we predict that BH-dragging can occur in more massive elliptical galaxies, and may be detected in future by improvements in observational techniques.

\clearpage

\appendix

\section{Trajectories of the SMBH remnant}
\label{sec:traj}
\begin{figure*}[b]
    \centering
    \includegraphics[width = 0.77\textwidth]{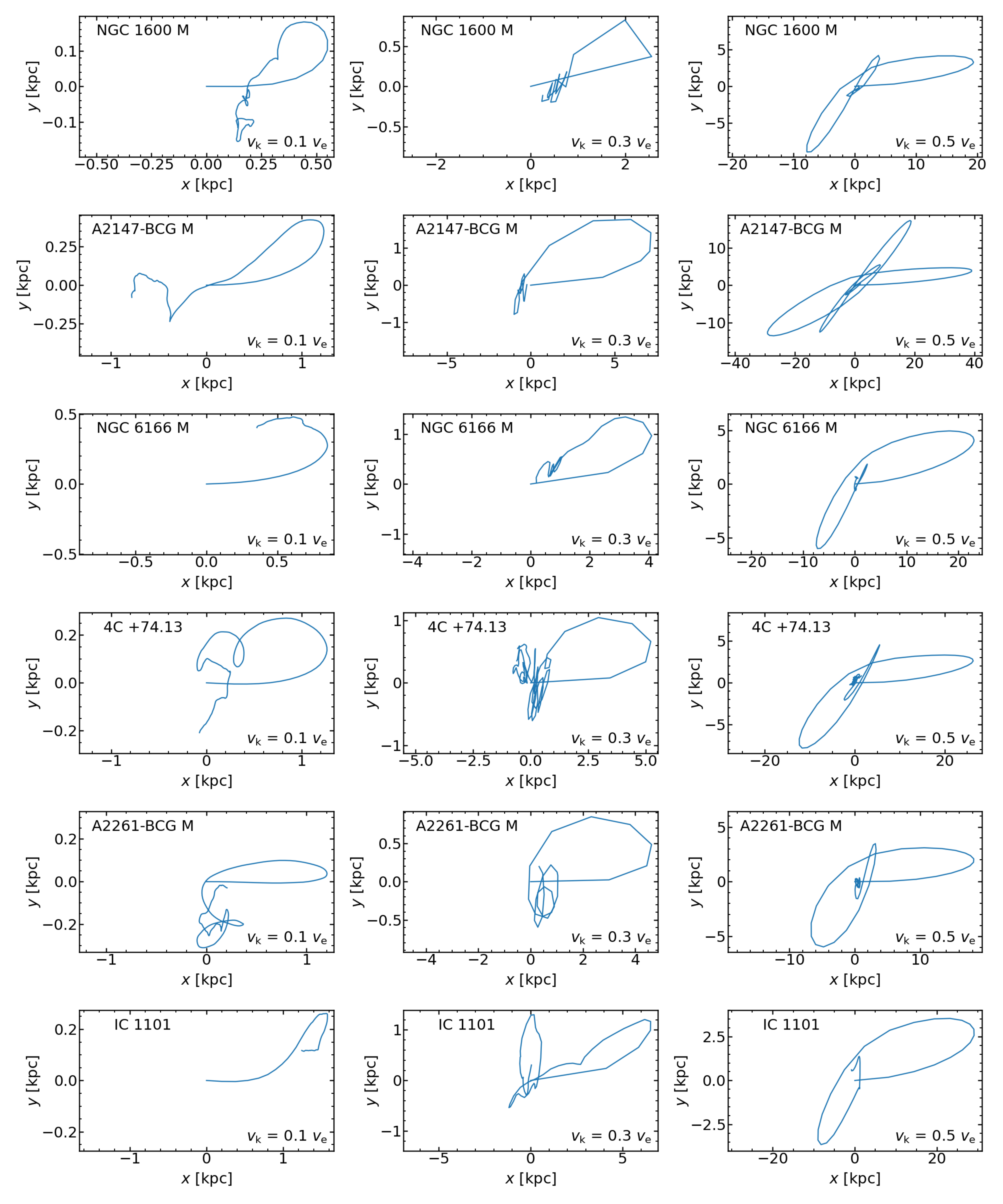}[b]
    \caption{Trajectory of the SMBH remnant (recentered on the COM) from the time the GW kick recoil is applied.
    Rows (from top to bottom) correspond to galaxies NGC 1600, A2147 BCG, NGC 6166, 4C +74.13,  A2261-BCG, IC 1101, respectively. Columns (from left to right) show results for $\vk / \ve$ of 0.1, 0.3 and 0.5, respectively. In contrast to the higher speed kicks in \citetalias{khonji2024core}, the SMBH in general makes just one or two passages through the nucleus before being damped to Brownian motion.}
    \label{fig:trajectory135}
\end{figure*}

Figure \ref{fig:trajectory135} shows the trajectory of the SMBH remnant after the GW recoil kicks. It generally makes only 1\mbox{--}2 passages before being quickly damped to Brownian motion.

\clearpage

\section{Density profiles}
\label{sec:density}

\begin{figure*}[h]
    \centering
    \includegraphics[width = \textwidth]{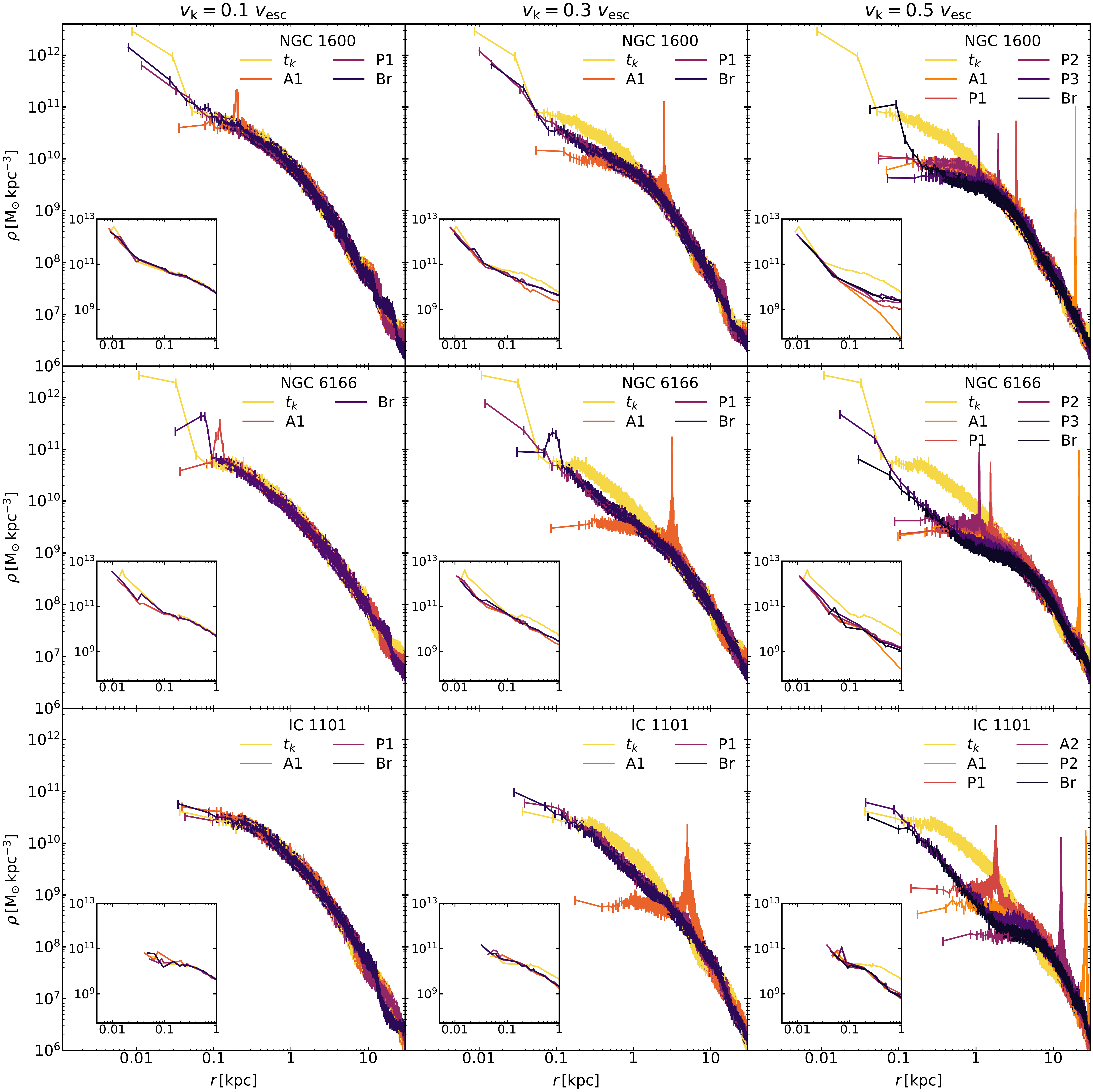}
    \caption{Volume density profiles, centered on the stellar COM, for the $N$-body merger remnant models of NGC 1600, NGC 6166 and IC 1101, using binned SPH profiles. The inset plots use spherical shells centered on the SMBH to show formation of the nucleus. Columns from left to right are for GW recoil kicks of $\vk/\ve=0.1$, $0.3$ and $0.5$ respectively. $\tk$ is the time when the $N$-body simulation is paused for the SMBH merger and the GW recoil kick is given; ``A$n$" and ``P$n$" indicate $n$th apocenter and pericenter passages of the kicked SMBH; ``Br" indicates that the SMBH has settled into Brownian motion. Increases in central density are seen at Br in almost all cases, with varying degrees of core formation outside them.}
    \label{fig:den1}
\end{figure*}

\begin{figure*}[h]
    \centering
    \includegraphics[width = \textwidth]{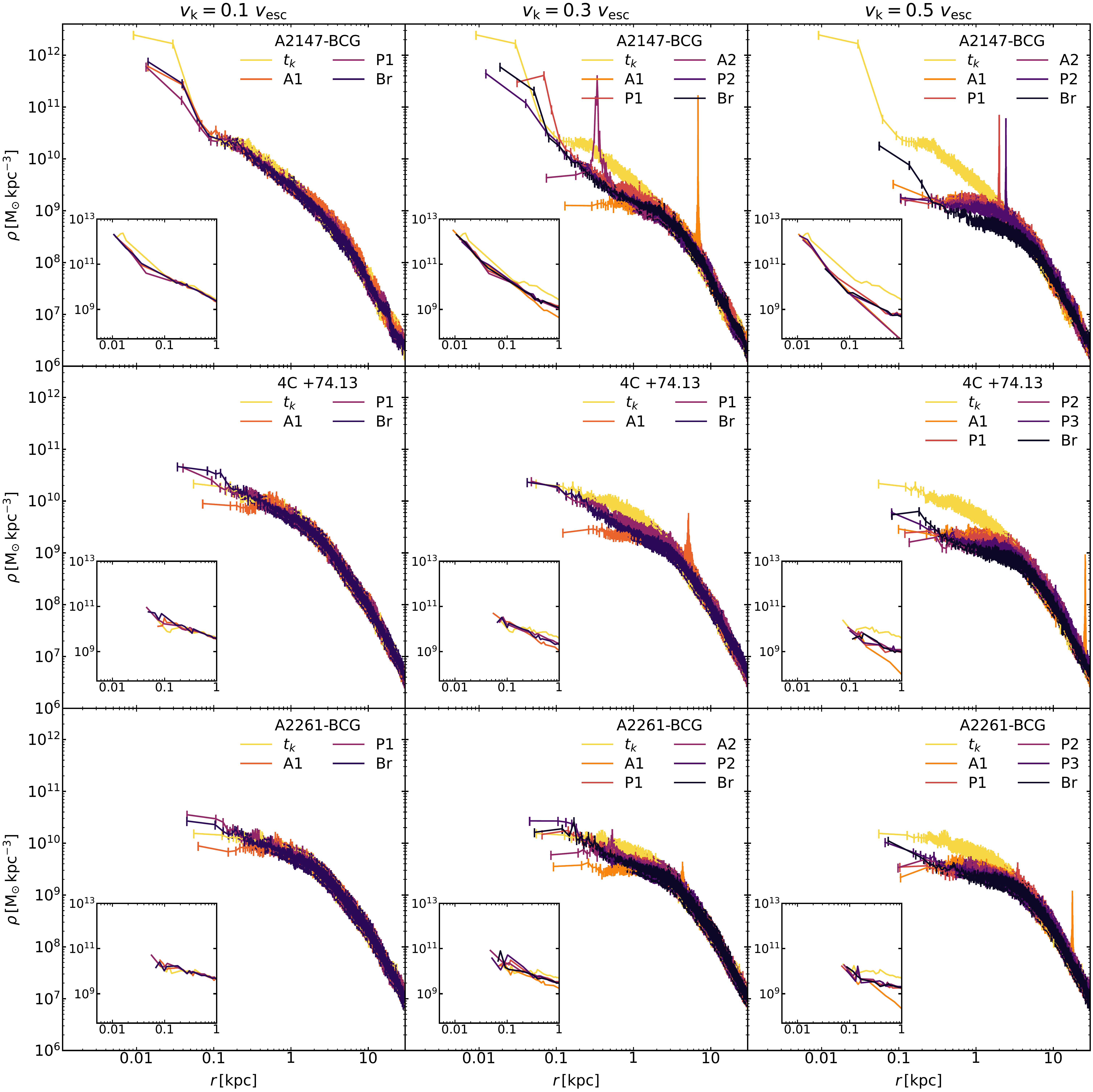}
    \caption{As for Figure \ref{fig:den1}, but for A2147-BCG, 4C +74.13 and A2261-BCG}
    \label{fig:den2}
\end{figure*}

\begin{figure*}
    \centering
    \includegraphics[width = 0.85\textwidth]{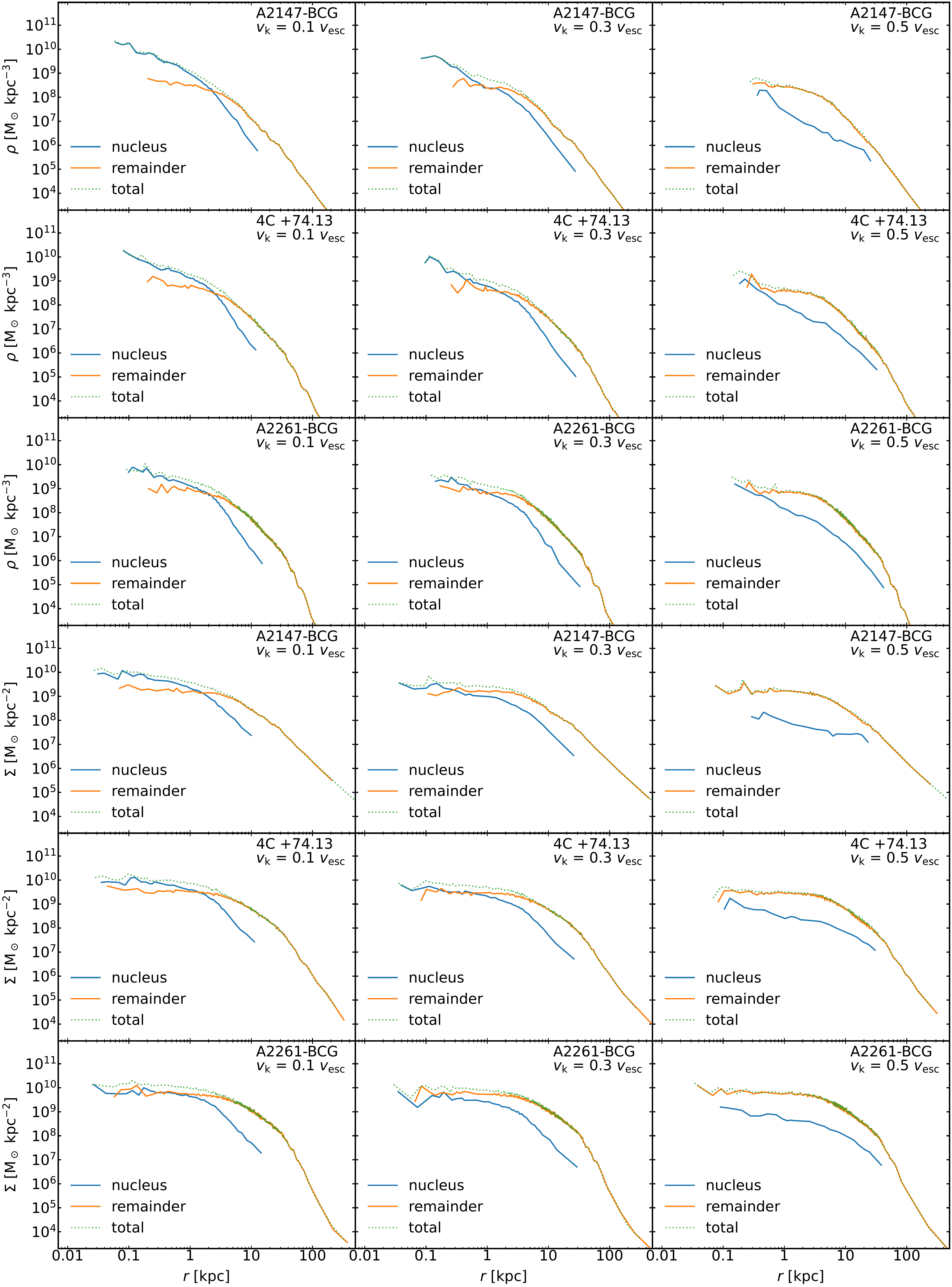}
    \caption{Volume (top three rows) and surface density (bottom three rows) profiles when the SMBH remnant has settled to Brownian motion for A2147-BCG, 4C +74.13, and A2261-BCG. Columns from left to right are for kicks with $\vk/\ve$ of $0.1$, $0.3$ and $0.5$ respectively. Profiles labeled ``nucleus" indicate those for particles that were within the influence radius of the SMBH at 1st apocenter (blue), and all other particles as ``remainder" (orange). The combined profiles are labeled ``total" (green dotted).}
    \label{fig:nuc_select2}
\end{figure*}

Figures \ref{fig:den1} and \ref{fig:den2} show the volume density profiles for all six galaxies, using a SPH-style kernel estimator \citep{pynbody}. Figure \ref{fig:nuc_select2} is as figure \ref{fig:nuc_select} but for galaxies A2147-BCG, 4C +74.13, and A2261-BCG.

\clearpage
\bibliography{nsc}{}
\bibliographystyle{aasjournal}

\end{document}